\documentclass[preprint,12pt]{elsarticle}

\usepackage[utf8]{inputenc}
\usepackage[english]{babel}

\usepackage{natbib}
\setcitestyle{authoryear,open={(},close={)}} %Citation-related commands

\usepackage{amssymb}
\usepackage{subfig}
\usepackage{url}
\usepackage{xcolor}
\usepackage{eqnarray,amsmath}
\usepackage{algorithm}
\usepackage{comment}
\usepackage{xcolor}
\usepackage{algpseudocode}
\usepackage{natbib}
\usepackage{etoolbox}

\journal{Journal of Statistical Planning and Inference}

\begin{document}

\begin{frontmatter}

%% Title, authors and addresses

%% use the tnoteref command within \title for footnotes;
%% use the tnotetext command for theassociated footnote;
%% use the fnref command within \author or \address for footnotes;
%% use the fntext command for theassociated footnote;
%% use the corref command within \author for corresponding author footnotes;
%% use the cortext command for theassociated footnote;
%% use the ead command for the email address,
%% and the form \ead[url] for the home page:
%% \title{Title\tnoteref{label1}}
%% \tnotetext[label1]{}
%% \author{Name\corref{cor1}\fnref{label2}}
%% \ead{email address}
%% \ead[url]{home page}
%% \fntext[label2]{}
%% \cortext[cor1]{}
%% \affiliation{organization={},
%%             addressline={},
%%             city={},
%%             postcode={},
%%             state={},
%%             country={}}
%% \fntext[label3]{}

\title{A Framework of Zero-Inflated Bayesian Negative Binomial Regression Models For Spatiotemporal Data}

%% use optional labels to link authors explicitly to addresses:
%% \author[label1,label2]{}
%% \affiliation[label1]{organization={},
%%             addressline={},
%%             city={},
%%             postcode={},
%%             state={},
%%             country={}}
%%
%% \affiliation[label2]{organization={},
%%             addressline={},
%%             city={},
%%             postcode={},
%%             state={},
%%             country={}}

\author[inst1]{Qing He}

\affiliation[inst1]{organization={Department of Statistics and Data Science},%Department and Organization
            addressline={
University of Central Florida}, 
            city={Orlando},
            postcode={32816}, 
            state={Florida},
            country={USA}}

\author[inst1]{Hsin-Hsiung Huang}

\begin{abstract}
Spatiotemporal data analysis with massive zeros is widely used in many areas such as epidemiology and public health. We use a Bayesian framework to fit zero-inflated negative binomial models and employ a set of latent variables from P\'olya-Gamma distributions to derive an efficient Gibbs sampler. The proposed model accommodates varying spatial and temporal random effects through Gaussian process priors, which have both the simplicity and flexibility in modeling nonlinear relationships through a covariance function. To conquer the computation bottleneck that GPs may suffer when the sample size is large, we adopt the nearest-neighbor GP approach that approximates the covariance matrix using local experts. For the simulation study, we adopt multiple settings with varying sizes of spatial locations to evaluate the performance of the proposed model such as spatial and temporal random effects estimation and compare the result to other methods. We also apply the proposed model to the COVID-19 death counts in the state of Florida, USA from 3/25/2020 through 7/29/2020 to examine relationships between social vulnerability and COVID-19 deaths.
\end{abstract}

%%Graphical abstract
%\begin{graphicalabstract}
%\includegraphics{grabs}
%\end{graphicalabstract}

%%Research highlights
%\begin{highlights}
%\item A novel model using Bayesian zero-inflated negative binomial (ZINB) regression with nearest-neighbor Gaussian processes (NNGP) for spatial and temporal random effects which are flexible for count responses with excess zeros.

%\item Nearest-neighbor Gaussian processes yield a lower dimensional parameter space and hence improve the estimation of spatial and temporal effects via much smaller matrix inversions of the spatial and temporal covariance matrices.

%\item Bayesian ZINB-NNGP leads to more scalable computation.

%\end{highlights}


\end{frontmatter}

%% \linenumbers

%% main text

\section{Introduction}
Zero-inflated models have been widely used for handling count data with excessive zeros \citep{lewsey2004utility, cheung2002zero}.
By construction, zero-inflated models assume that the zeros come from one of the cases \citep{greene1994accounting}: 1) structural zeros corresponding to individuals who are not at risk for an event, and therefore have no opportunity for a positive count and 2) random zeros which correspond to a latent class of individuals who are at risk for an event but nevertheless have an observed response of zero.  
A zero-inflated negative binomial (ZINB) model is a popular choice for modeling zero-inflated data as it gives more reliable parameter estimates when the nonzero counts are over-dispersed compared to other models like the zero-inflated Poisson model \citep{yau2003zero}.
Bayesian approaches to fitting zero-inflated models have gained attention recently \citep{ghosh2006bayesian,neelon2010bayesian,zuur2012zero}. There is an abundance of advantages of Bayesian zero-inflated models including using prior beliefs with Bayesian inference instead of deriving asymptotically approximate distributions for estimation of parameters \citep{gelman1995bayesian}. Bayesian approaches produce tractable inference for ZINB models and have been implemented through software such as R \citep{R}, Stan \citep{Stan}, and WinBUGS \citep{lunn2013bugs}.

It is important to account for spatial and temporal structures in areas such as epidemiology and public health. Within the Bayesian framework, it is natural to build a Bayesian hierarchical model with a prior distribution for spatial and temporal effects \citep{kang2011bayesian}. Distance-based exponential or Mat\'{e}rn covariance functions are commonly adopted to specify spatial correlations in geostatistical data, while conditionally autoregressive models are often used for temporal effects \citep{cressie2015statistics,banerjee2003hierarchical,zhang2022application} through a hierarchical modeling framework \citep{cressie2015statistics}.
\cite{neelon2022multivariate} proposed a cubic B-spline model to fit both temporal (overall time trend) and spatial (county-specific trend) random effects in a Bayesian framework, which depends on the number and locations of the interior knots and specification of the degree of basis functions.

There is limited research with the focus on the Bayesian ZINB model with the consideration of flexible spatial and temporal effects \citep{wang2016dynamic, gu2020modeling}. \cite{neelon2019bayesian} proposed a Bayesian ZINB model with spatial and temporal random effects but assumed that temporal effects changed in a fixed increment, which may not reflect the randomness nature of temporal effects. In this study, we focus on proposing a flexible approach to model spatial and temporal effects in a Bayesian ZINB framework. A promising option is using a Gaussian process (GP) prior which characterize spatial or temporal effects through a kernel function and enables tractable nonparametric Bayesian inference. A GP prior is a probability distribution over infinite number of possible functions which make it a powerful tool to model nonlinear patterns. GP accounts for quantification of uncertainty, which relies on the predictive conditional distribution of a new spatial location given a set of observed locations \citep{cousin2016kriging}. GP can be viewed as a spline in a reproducing kernel Hilbert space with the reproducing kernel of a covariance function \citep{kimeldorf1970correspondence,wahba1990spline}. It has become a popular modeling tool in multivariate and geostatistical settings. For example, \citet{diana2021fast} proposed a Bayesian hierarchical occupancy model using GPs within a logistic regression framework for spatial and temporal random effects.
However, the covariance matrix used in GP models may hinder its implementation for large datasets due to computational issues. Recent research has developed various approximation techniques to overcome the computational limitations \citep{smola2001sparse, quinonero2005unifying, snelson2007local,  lazaro2010sparse}. There are two broad classes of these techniques, i.e., global approximation methods and local approximation methods \citep{liu2020gaussian}. For global approximation methods, sparsity of the covariance matrix is achieved through global distillation including using a subset of the training data, exploiting the sparse structure of the matrix, and low-rank models \citep{chalupka2013framework, quinonero2005unifying, titsias2009variational}. For example, low-rank models embed the original process into a lower-dimensional subspace. However, it usually requires a large subspace when used for large spatial datasets. Local approximation methods are based on multiple sets of local experts to improve the scalability based on the concept of divide-and-conquer \citep{liu2020gaussian, gramacy2016lagp}. This class of approximations has the capability of capturing nonstationary features from the localized experts. Depending on the partition of the input space, there are inductive local experts which employs static partitioning \citep{vasudevan2009gaussian} and transductive local experts which employ a dynamic partition \citep{datta2016hierarchical,datta2016nonseparable}.
As an example of transductive local experts, \citet{datta2016hierarchical} proposed a nearest-neighbor Gaussian process (NNGP) that uses conditional independence given information from neighboring points for large geostatistical datasets. It has well-defined sparse precision matrices for its finite-dimensional realization and provides fully process-based inference on underlying spatial processes.

In this paper, we adopt the NNGP model \citep{datta2016hierarchical} to model the spatial and temporal effects because of its ease of use and well-defined sparse precision matrix. The proposed Bayesian zero-inflated negative binomial (ZINB) model uses nearest-neighbor Gaussian process priors to fit spatial and temporal random effects that account for the over-dispersed and zero-inflated count response by incorporating the spatial and temporal covariance matrices. It allows us to obtain more reliable and stable inferences since NNGP yields a lower-dimensional spatial and temporal covariance matrices and parameter space, therefore requires much smaller matrix inversions.  
The propose framework is a scalable hierarchical ZINB model with flexible spatial and temporal effects through GP. First, the hierarchical model is based on work in \cite{neelon2019bayesian} to estimate regression parameters and the spatial and temporal effects. A key component to the parameter estimator is to introduce a set of latent variables through the P\'{o}lya-Gamma (PG) data augmentation, an underlying latent variable methodology \citep{pillow2012fully, polson2013bayesian}. With the PG scheme, posterior inference is highly efficient via Gibbs sampling by updating from conditional distributions with conjugate priors and explicit conditional normal distributions. 
To illustrate Gibbs sampling for generating a sequence of samples, we use an example of a pair of random variables $(\mathcal{X},\mathcal{Y})$.  A sequence of random variables $$\mathcal{X}'_0,\mathcal{Y}'_0,\mathcal{X}'_1,\mathcal{Y}'_1,\ldots,\mathcal{X}'_t,\mathcal{Y}'_t,\ldots,$$ is generated iteratively as follows: 1. set $\mathcal{X}'_0$ at some initial value; 2. sample the rest from $\mathcal{Y}'_j\sim f(y\mid \mathcal{X}'_j=x'_j)$ and $\mathcal{X}'_{j+1}\sim f(x\mid \mathcal{Y}'_j=y'_j)$ alternately \citep{casella1992explaining}.
It has been shown that the PG method yields superior performance compared to other Bayesian methods in the context of structural equation models with logistic regression for binary variables \citep{kim2018improved} and the Bayesian ZINB \citep{neelon2019bayesian}. 
Therefore we propose a more flexible structure for the spatial and temporal random effects through the GP prior in order to achieve fully model-based Bayesian inference for the spatial and temporal effects. More specifically, to facilitate the computation process, we incorporate the framework with a conjugate latent NNGP model \citep{datta2016hierarchical,datta2016nonseparable,zhang2021high}, which is scalable for massive spatial datasets on modest computing environments \citep{datta2016hierarchical,datta2016nonseparable, zhang2021high}.

\section{Bayesian Inference for ZINB Model Using NNPG}
\subsection{The P\'{o}lya-Gamma Scheme}

Let $y_i$ as the outcome of the $i$th observation, which may be a count $y_i = 0, 1, 2, \ldots$ or a binary indicator $y_i = 0, 1$.
We aim to model $Y_i$ and a set of covariates $X_i = [1, x_{i1}, \ldots , x_{iP} ]^T$ through
$
\mbox{E}[Y_i\mid X_i] = g^{-1}
(X_i\beta),
$
where $\beta= [\beta_0, \ldots , \beta_P ]^T$ are the
regression coefficients and $g$ is the canonical link function in generalized linear models (GLMs) \citep{mccullagh2019generalized}.
We utilize the efficient PG scheme for estimating the regression coefficients in GLMs. An advantage of the PG scheme is that the posterior distributions of the parameters of interest enable us to
make use of Gibbs sampling approach within logistic regression and NB regression frameworks \citep{polson2013bayesian} through an augmented PG distributed random variable $\omega$.
Generally, assuming a prior distribution for regression coefficient $\beta\sim \mathcal{N}(b,\Sigma_0)$, the full conditional distributions induced by PG for the Gibbs sampler are:
\begin{align*}
    \omega_i\mid\beta \sim \mathcal{PG}(n_i,X_i\beta), \mbox{ and }
    \beta \mid \Omega, z \sim \mathcal{N}(\mu, \Sigma),
\end{align*}
where
\begin{align*}
  \Sigma =
(\Sigma_0^{-1} +X^{T}
\Omega X)^{-1}, \mbox{ and }
\mu = \Sigma
\left(\Sigma_0^{-1}b +X^{T}
\Omega z\right).
\end{align*}
Here $\mathcal{PG}$ denotes a PG distribution, $ \Omega = \text{diag}(\omega_1,\ldots, \omega_n)$ and $z$ is the latent variable with $z_i =\frac{y_i -n_i/2}{\omega_i}$.
Specifically, we use the PG scheme for estimating the regression coefficients in the Bernoulli and NB parts of the ZINB regression model, respectively.  

In the binomial logistic regression model (logit model), let $n_i=1\;\forall i$ and $y_i \sim \text{Binom} (1, \frac{1}{1+e^{-\eta_i}})$  where $\eta_i = x_i^T \beta$ .
Following \cite{polson2013bayesian}, the full conditional for $\beta$ in a binomial regression is
$$
p(\beta\mid y,r,\omega) \propto \pi(\beta)\exp\left[-\frac12
(z -X\beta)^T \Omega (z -X\beta)\right],    
$$
where $\pi(\beta)$ is the prior distribution, 
$z$ is an $n \times 1$ vector with
$z_i =\frac{y_i -1/2}{\omega_i}$ and $\Omega =$ diag$(\omega_i)$. It is readily seen that $z\mid \beta, \Omega \sim \mathcal{N}(X\beta, \Omega)$, which leads to the convenient Gibbs sampler for $\beta$.
In the negative binomial regression model \citep{pillow2012fully}, $n_i=y_i+r$ and $r$ can be obtained using either the Metropolis–Hastings algorithm in which a uniform prior is applied with positive candidate values of $r$ are drawn from a zero-truncated normal proposal centered at the current value of $r$, or the two-stage Gibbs sampling 
\citep{zhou2013negative,dadaneh2018bayesian}.
Consider the following model for a count response $Y_i$,
\begin{align*}
  p(Y_i=y_i \mid r,\beta) &\stackrel{d}{=}\frac{\Gamma(y_i + r )}{\Gamma(r )y_i !}
(1-\psi_i )^{r}(\psi_i )^{y_i},\; r > 0,\;\mbox{ where }\\
\psi_i &=\frac{\exp(x^T_i\beta)}{1 + \exp(x^T_i\beta)} =
\frac{\exp(\eta_i)}{1 + \exp(\eta_i)},
\end{align*}
where the NB probability parameter $\psi_i$ is parameterized
using the expit (inverse-logit) function,
which allows us to apply the same properties of the P\'{o}lya-Gamma
density as in the logistic case.

 The mean and variance of $Y_i$ are
 \begin{align*}
   \mbox{E}(Y_i \mid r ,\beta) &=\frac{\psi_i}{1-\psi_i}=r\exp(\eta_i)=\mu_i,\\
\mbox{Var}(Y_i \mid r ,\beta) &=
 \frac{r\psi_i}{(1-\psi_i)^2}= r \exp(\eta_i ) [1 + \exp(\eta_i )]
 = \mu_i (1 + \mu_i/r ) .
 \end{align*}
 The parameter $r$ accounts for overdispersion in the data.
 As $r \to \infty$, the counts become increasingly dispersed and the NB distribution converges to the Poisson distribution.
The above parameterization also leads to a Gaussian linear model where 
$
z\mid\beta, \Omega \sim \mathcal{N}(X\beta, \Omega)
$
where
$z$ is an $n \times 1$ vector with
$z_i =\frac{y_i -r}{2\omega_i}$, $\omega_i \sim \mathcal{PG}(y_i+r,X_i\beta)$ and $\Omega =$ diag$(\omega_i)$.

\section{Zero-inflated Negative Binomial Model}

\subsection{The Proposed Spatiotemporal ZINB-NNGP Model}

We are interested in a spatiotemporal random intercept model on data with excessive zeros. Assume observations are collected at a fixed collection of distinct locations denoted by $\mathbf{S} = \{1,\ldots,S\}$ and across $T$ distinct time segments (e.g., weeks, months, seasons and years) denoted by $\mathbf{T} = \{1,\ldots,T\}$. We index the observations by $j = 1, \ldots N$, where $N$ is the total sample size. For a sampling unit $\{s,t\}$ where $s \in \mathbf{S}$ and $t \in \mathbf{T}$, we denote its number of observations by $n_{s,t}$ ($n_{s,t}\geq 0$). In a special case, $n_{s,t}=1, \forall s \in \mathbf{S}, t \in \mathbf{T}$, that is there is exactly one observation for each sampling unit $\{s,t\}$ and $N = S \times T$. In the proposed framework, it does not require the sampling units to have equal sample sizes or at least one observation. Sparsity and unequal sizes are considered in the simulation study.

Let response $Y_{j}$ be the observed count of the $j$th observation, $\phi_j$ be the probability of belonging to the at-risk group, and $\psi_j$ be the success rate for observations in the at-risk group. The Bayesian ZINB model \citep{neelon2019bayesian} with a latent at-risk indicator variable $W_{j}$ is introduced as follows:
$$
Y_{j} \sim (1 -\phi_{j})1_{(W_{j}=0\land Y_{j}=0)} + \phi_{j}\mathcal{NB}(\mu_j , r )1_{(W_{j}=1)},\; j = 1,\ldots, N,
$$
where $\mu_j$ is the mean of the NB distribution and $r$ is the overdispersion parameter. A value of $1$ for $W_j$ means that the $j$th observation belongs to the at-risk group, and a value of $0$ means the ``not at-risk" group. Let $s_j$ and $t_j$ be its spatial location and time point, respectively. We model the at-risk probability $\phi_j$ using a logit model and $Y_j \mid W_j = 1$ using a negative binomial regression model: 
\begin{align}
    Pr(Y_{j} = 0) &= (1-\phi_{j} ) + \phi_{j} (1 -\psi_{j} )^r, \nonumber \\
\mbox{logit}(\phi_{j} )
&= \mbox{logit} [Pr(W_{j} = 1\mid \alpha,a_{s_{j}},b_{t_{j}},\epsilon_{11s_{j}},\epsilon_{12t_{j}})] \nonumber\\
&= x^T_{j} \alpha +a_{s_{j}}+b_{t_{j}}+\epsilon_{11s_{j}}+\epsilon_{11t_{j}} \nonumber\\
 &= \eta_{1j},\\
p(y_{j} \mid r , \beta,c_{s_{j}},d_{t_{j}},\epsilon_{21s_{j}},\epsilon_{22t_{j}},W_{j} = 1)&\stackrel{d}{=} 
\frac{\Gamma(y_{j} + r )}{\Gamma(r )y_{j} !}
(1 -\psi_{j} )^r \psi_{j}^{y_{j}},\; \forall j\;\mbox{ s.t. }W_{j} = 1,\;\nonumber \\
 \psi_{j} &=\frac{\exp(x^T_{j}\beta + c_{s_{j}}+d_{t_{j}}+\epsilon_{21s_{j}}+ \epsilon_{22t_{j}})}{1+\exp(x^T_{j}\beta+ c_{s_{j}}+d_{t_{j}}+\epsilon_{21s_{j}}+\epsilon_{22t_{j}})} \nonumber\\
 &=\frac{\exp(\eta_{2j})}{1+\exp(\eta_{2j})},
\label{ZINB model}
\end{align}
where $a_{s_j}$ and $\epsilon_{11{s_j}}$ are the spatial random effects and spatial random noise in the logit model, $b_{t_j}$ and $\epsilon_{12{t_j}}$ are the temporal random effects and temporal random noise in the logit model, $c_{s_j}$ and $\epsilon_{21{s_j}}$ are the spatial random effects and spatial random noise in the negative binomial regression model, and $d_t$ and $\epsilon_{22{t_j}}$ are the temporal random effects and temporal random noise in the negative binomial model. The two models are often referred as the binary component (the logit model) and count component (the negative binomial model), which we will use extensively in the paper. The proposed model incorporates additive spatial and temporal random effects as intercepts in the binary and count component. Compared with the existing Bayesian ZINB model  \citep{neelon2019bayesian}, the proposed model is more flexible through the use of GP priors on the spatial and temporal random effects.
To elaborate the GP priors, first we define the spatial random effects corresponding to the sampling locations $\mathbf{S}=\{1,\ldots,S\}$ as 
\begin{align*}
\vec{a} &= (a_1,\ldots, a_{S})^T, \mbox{in the binary component and }\\
\vec{c} &= (c_1,\ldots, c_{S})^T, \mbox{in the count component}.
\end{align*}
Similarly, we define the temporal random effects of temporal points $\mathbf{T}=\{1,\ldots,T\}$ as
\begin{align*}
\vec{b} &= (b_1,\ldots, b_{T})^T, \mbox{in the binary component and}\\
\vec{d} &= (d_1,\ldots, d_{T})^T, \mbox{in the count component}.
\end{align*}

A list of descriptions of the notations is given in Table~\ref{notation desc}.
\begin{table}[H]
\footnotesize
\begin{tabular}{ p{3cm}|p{8.8cm}}
 \hline
%  \multicolumn{4}{|c|}{City 4 (Testing dataset):V16-Method 2} \\
%  \hline
Notation & Description \\
\hline
$S$ & the number of spatial locations\\
$T$ & the number of temporal points\\
$\mathbf{S} = \{1,2,\ldots, S\}$ & the list of spatial locations\\
$\mathbf{T} = \{1,2,\ldots, T\}$ & the list of temporal points\\
$N$ & the number of observations\\
$Y_j, j =1,\ldots,N$ & the response of the $j$th observation\\
$s_j \in \mathbf{S} $ & the spatial location of the $j$th observation\\
$t_j \in \mathbf{T}$ & the temporal point of the $j$th observation\\
$\phi_j$ & the probability of belonging to the at-risk group\\
$\psi_j$ & the success rate for observations in the at-risk group\\
$W_j$ & the indicator for belonging to the at-risk group\\
$\mu_j$ & the mean of negative binomial distribution\\
$r$ & the dispersion parameter of negative binomial distribution\\
$a_{s_j}$ & the spatial random effect in the logit model\\
$\epsilon_{11s_j}$ & the spatial random noise in the logit model\\
$b_{t_j}$ & the temporal random effect in the logit model\\
$\epsilon_{12t_j}$ & the temporal random noise in the logit model\\
$c_{s_j}$ & the spatial random effect in the negative binomial model\\
$\epsilon_{21s_j}$ & the spatial random noise in the negative binomial model\\
$d_{t_j}$ & the temporal random effect in the negative binomial model\\
$\epsilon_{22t_j}$ & the temporal random noise in the negative binomial model\\
\hline
\end{tabular}
% \captionof{table}
\caption{The description of the notations used in the proposed model.}
\label{notation desc}
\end{table}

To account for correlations between spatial locations or temporal points, we employ GPs for both the spatial and temporal effects. Specifically, we assume that $\vec{a} = (a_1, \ldots, a_S )^T$ in the binary component is distributed according to a GP with parameters $(\sigma_{11},l_{11} )$. Let $h_s$ be the specific location information of site $s$ (e.g., latitude and longitude) which can be used to compute distance between different locations. It corresponds to assuming that $(a_1, \ldots, a_S )^T \sim \mathcal{N}(0,K_{ \sigma_{11},l_{11}} (h_1, \ldots, h_S ))$,
where $K_{\sigma_{11},l_{11}} (h_1, \ldots ,h_S )_{i,j} = \sigma_{11}^2e^{-\frac{|h_i-h_j|^2}{l_{11}^2}}, i,j=1,\ldots,S$. Parameter $\sigma_{11}$ determines the overall variability of the GP, while parameter $l_{11}$ measures the correlation between location $i$ and $j$. Similarly, the autocorrelated temporal random effect $\vec{b} = (b_1, \ldots , b_T)^T$ in the count component is
distributed according to a GP with parameters $(\sigma_{12},l_{12})$. Let $w_t$ be the index of temporal point $t$. For example, if the data are at monthly level spanning from January to December, then the corresponding indices could be $(1,\ldots, 12)$. The GP prior can be written as $(b_1, \ldots ,b_T)^T \sim \mathcal{N}(0,K_{\sigma_{12},l_{12}} (w_1, \ldots, w_T))$.

Let $V_1$ and $V_2$ be the $N\times S$ and $N\times T$ design matrices for the spatial and temporal random effects for the observed data. The following hierarchical structure completes the definition of the binary component with spatial random effect $\vec{a}=(a_1,\ldots,a_S)^T$, temporal random effect $\vec{b}=(b_1,\ldots,b_T)^T$, spatial random noise
$\vec{\epsilon}_{11}=(\epsilon_{111},\ldots,\epsilon_{11S})^T$, and temporal random noise $\vec{\epsilon}_{12}=(\epsilon_{121},\ldots,\epsilon_{12T})^T$ including the prior distributions of relevant parameters.
The hierarchical structure for the negative binomial regression with random effects $\vec{c}=(c_1,\ldots,c_S)^T$ , $\vec{d}=(d_1,\ldots,d_T)^T$,
$\vec{\epsilon}_{21}=(\epsilon_{211},\ldots,\epsilon_{21S})^T$, and $\vec{\epsilon}_{22}=(\epsilon_{221},\ldots,\epsilon_{22T})^T$  are in the same fashion.
\begin{align}
(a_1, \ldots , a_S)^T &\sim \mathcal{N}(0,K_{\sigma_{11},l_{11}} (h_1, \ldots , h_S)),\nonumber\\
\sigma^2_{11} &\sim \mathcal{IG}(a_{\sigma_1} , b_{\sigma_1}),\;
l_{11} \sim Gamma(a_{l_1} , b_{l_1}),\nonumber\\
(b_1, \ldots, b_T )^T &\sim \mathcal{N}(0,K_{\sigma_{12},l_{12}} (w_1, \ldots ,w_T )),\nonumber\\
\sigma^2_{12} &\sim \mathcal{IG}(a_{\sigma_2} , b_{\sigma_2}), \;
l_{12} \sim Gamma(a_{l_2} , b_{l_2}),\nonumber \\
\epsilon_{11s} & \stackrel{iid}\sim \mathcal{N}(0, \sigma^2_{\epsilon_{11}}),\; \sigma^2_{\epsilon_{11}} \sim \mathcal{IG}(a_{\epsilon}, b_{\epsilon}), s = 1, \ldots, S,\nonumber \\
\epsilon_{12t} & \stackrel{iid}\sim \mathcal{N}(0, \sigma^2_{\epsilon_{12}}),\; \sigma^2_{\epsilon_{12}} \sim \mathcal{IG}(a_{\epsilon}, b_{\epsilon}), t = 1, \ldots, T.\nonumber
\end{align}

\subsubsection{Latent Nearest-Neighbor Gaussian Process}\label{NNGP section}
% NNGP is
% derived from the conditional specification of the joint distribution of the spatial random effects  \citep{datta2016hierarchical,datta2016nonseparable} in

In a spatial regression model at location $s \in \mathbf{S}$ and time $t \in \mathbf{T}$ in a spatiotemporal domain $\mathcal{D}$, assume that
\begin{align*}
    y(s,t) = \mu + w_1(s) + w_2(t)+ \epsilon_1(s)+\epsilon_2(t),\; (s,t) \in \mathcal{D},
%\label{latent GP example}
\end{align*}
where $\mu$ is the mean function that is itself modeled as a linear combination of known
covariates, $w_1(s)$ follows a latent GP with mean zero and a positive-definite cross-covariance matrix $C_\phi(s,s')$,
% $\sigma^2 \rho_\phi\left(s, s'\right)$,
and $\epsilon_1(s)$ follows a normal distribution with mean zero and variance $\sigma^2$. In the same fashion, we can define the temporal random effect $w_2(t)$ and the temporal random noise $\epsilon_2(t)$. This type of regression model is called a latent GP model since the GP is used as a prior in the latent process $w(s)$ and $w(t)$, while a response GP model imposes a GP on the outcome $y(s,t)$. 
In the proposed Bayesian spatiotemporal ZINB model \eqref{ZINB model}, we use the latent GP model with both spatial and temporal random effects.

In NNGP \citep{datta2016nonseparable}, the joint density of $(w_1, \cdots , w_S)$ was approximated by a series of conditional densities of size at most $m$, where $m \leq S$ and $S$ is the size of distinct spatial locations. It was shown that the approximation was essentially a multivariate Gaussian density with covariance matrix $\tilde{C}_\phi(\mathbf{S},\mathbf{S})$. Using the spatial random effects $w_1(\mathbf{S})$ described above as an example, 
% let $M= \rho_\phi+I$.The covariance function can be re-written as $C=\sigma^2 M$.
by selecting a list of nearest neighbors with size $m$ for each location in $\mathbf{S}$, the Gaussian density $p(w_1(\mathbf{S}))$ can be approximated as follows \citep{datta2016nonseparable, zhang2021high}:
\begin{align}
p(w_1(\mathbf{S}))=N(w_1(\mathbf{S})\mid \mathbf{0},C_\phi(\mathbf{S},\mathbf{S})) \approx N(w_1(\mathbf{S})\mid \mathbf{0},\tilde{C}_\phi(\mathbf{S},\mathbf{S})).
\label{density approximation}
\end{align}
The approximation $\tilde{C}_\phi(\mathbf{S},\mathbf{S})$ is computationally efficient because its inverse $\tilde{C}_\phi(\mathbf{S},\mathbf{S})^{-1}$ is sparse and can be written as:
$$
\tilde{C}_\phi(\mathbf{S},\mathbf{S})^{-1}=(I-A_{S})^T D^{-1}_{S} (I-A_{S}),
$$
where $A_{S}$ is a lower triangular matrix and $D_{S}$ is a diagonal matrix. To compute ${A}_{S}$, suppose $N_m(i)$ be the set of column indices of the $m$ nearest neighbors that contain nonzero entries in the $i$th row of ${A}_{S}$, where $m=1,\ldots, S-1$ is a hyperparameter controlling the size of nearest neighbors. Let ${A}_{S} = [{a}_1 : \cdots : {a}_S]^T $
and $D_{S} = \text{diag}(d_1, d_2, \ldots , d_S)$. The first row of ${A}_S$ has all elements equal to 0 and $d_1 = C_\phi(1,1)$. For $i = 2, \ldots, S$, we obtain the nonzero entries at column positions indexed by $N_m(i)$ in $A_S$ and the diagonal elements in $D_S$ as follows:
\begin{align}
% \begin{split}
a_i(N_m(i)) &= C_\phi(i, N_m(i)) C_\phi(N_m(i), N_m(i))^{-1} \mbox{ and }
\label{constrcut A}
\end{align}
\begin{align}
d_i &= C_\phi(i, i) - 
C_\phi(i, N_m(i))
C_\phi(N_m(i), N_m(i))^{-1}C_\phi(N_m(i), i).
% \end{split}
\label{constrcut D}
\end{align}
\vspace{-.5cm}
\subsubsection{Spatio-temporal correlations via NNGP in ZINB} \label{section: NNGP}
In this section, we describe how to embed spatiotemporal correlations through NNGP in the proposed Bayesian ZINB framework. As we assume GP priors for both the spatial and temporal random effects in the proposed model, the latent NNGP can be used for one random process or both when needed. Since the latent NNGP is designed for a spatial process initially, we use the spatial random effects to demonstrate the application of latent NNGP for easier understanding. When the number of spatial locations $S$ is large, the computation bottleneck in the conjugate Gaussian regression model lies in computing $K_{\sigma_{11},l_{11}}^{-1}$ and $K_{\sigma_{21},l_{21}}^{-1}$ when updating $\sigma_{11}$ and $\sigma_{21}$. To facilitate it, we can find a sparse alternative for the Cholesky decomposition of $K_{\sigma_{11},l_{11}}^{-1}$ and $K_{\sigma_{21},l_{21}}^{-1}$. 
We use the covariance matrix for the spatial random effects from the binary component as an example to illustrate the methodology. 
First, we rewrite the covariance matrix $C_\phi(\mathbf{S},\mathbf{S}) = K_{\sigma_{11},l_{11}}(h_1, \ldots , h_S)=\sigma_{11}\rho_{l_{11}}$. Let $\tilde{\rho}_{l_{11}}$ be the NNGP approximation of $\rho_{l_{11}}$ with its inverse $\tilde{\rho}^{-1}_{l_{11}}=(I-A_S)^T D^{-1}_S (I-A_S)$, as described in Section \ref{NNGP section}. 
% That is $M$ in \eqref{density approximation} becomes $\rho_\phi$.
Nonzero entries in the $i$th row where $i=2,\ldots, N$ of $A_S$ and $D_S$  in \eqref{constrcut A} and \eqref{constrcut D} become
\begin{align}
% \begin{split}
a_i(N_m(i)) &= \rho_{l_{11}} (i, N_m(i)) (\rho_{l_{11}}(N_m(i), N_m(i)))^{-1}
\label{constrcut A_new}
\end{align}
and
\begin{align}
d_i &= 1 - 
a_i(N_m(i))
\rho_{l_{11}}(N_m(s_i), i).
% \end{split}
\label{constrcut D_new}
\end{align}

In this same fashion, we model the temporal correlation. The latent NNGP is applied to facilitate to computation bottleneck when updating the hyperparameters for the spatial and temporal random effects. Hence, when the spatial size and/or temporal size are large, the latent NNGP can be used. Otherwise, one may just compute the inverse of the covariance matrix. The step-by-step posterior sampling procedure is described in the appendix.

\section{Simulated and Real Data Analyses}
In this section, we apply the proposed Bayesian ZINB-NNGP model to both simulated and COVID-19 fatality data.
For the simulation data, a total number of  $N$ observations are generated from $S$ locations across $T$ time points. First, we consider exactly one repetition for each sampling unit $(s,t)$, we vary the spatial sizes to show if the proposed model can be generalized well to a large spatial dimension. We set $S=200$ and $T=20$ in simulation 1 and set $S=500$ and $T=20$ in simulation 2. Second, we relax the constraint on the number of repetitions for each sampling unit by simulating a dynamic number of repetitions for each sampling unit from a Poisson distribution. This is shown in simulation 3.

For the COVID-19 analysis, the daily level data on the total number of COVID-19 deaths in Florida counties in the early stage from 3/25/2020 through 7/29/2020 are analyzed. There are 69 counties ($S=69$) and 127 days ($T=127$) in the dataset. The reported death count of county $s$ at day $t$, denoted by $y_{s,t}$, is the response variable. We apply the proposed model to analyze the association of COVID-19 death at-risk and count rates with social vulnerability, other sociodemographic characteristics (i.e., health insurance coverage, urbanicity), population health care resources (i.e., primary care physicians), population health measures (i.e., life expectancy, obesity), and population density.

\subsection{Simulation Data}
\subsubsection{Simulation 1: a moderate spatial dimension and one repetition in each sampling unit}

In simulation~1, we have simulated a dataset with $S=200$ and $T = 20$, which results in a total sample size of $N=4,000$. For the fixed effect, we sample $x\sim\mathcal{N}(0,1)$, set the intercept and slope in the binary component at $\alpha_0=-0.25$ and $\alpha_1=0.25$, and the intercept and slope in the count component as $\beta_0=0.5$ and $\beta_1=-0.25$. The spatial locations are generated from a unit square. The latent NNGP approximation is used when updating $\sigma_{11}$ and $\sigma_{21}$ for the spatial random effects. For the size of nearest neighbors used in the latent NNGP model, we have tested different values from 8 to 15 and found that the results are relatively stable. In the results reported for both the simulations and COVID analysis, we set the size of nearest neighbors as 13. The posterior means and 95\% credible intervals as well as the true values of the parameters and hyperparameters are presented in Table~\ref{res:simulation1}. We also use the Bayesian spatiotemporal ZINB model with random effects proposed by \citet{neelon2019bayesian} and ZINB regression models from R package `pscl' \citep{jackman2015package} for comparison. 
Due to the difference in model structure, both comparative models do not have results for the hyperparameters used in the proposed framework.

\begin{figure}[H]
\begin{center}
\includegraphics[scale=0.6]{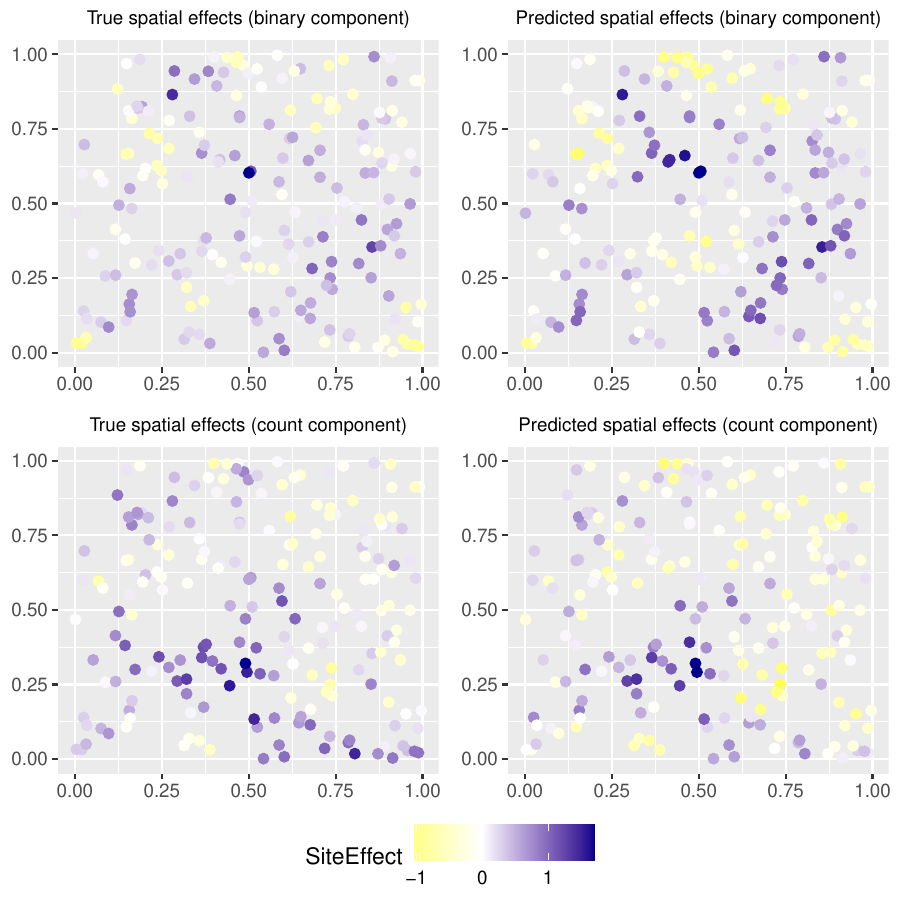}
    \caption{The true and fitted spatial effects in simulation 1 using the proposed model. Each dot represents a location and the color represent the scale of the spatial random effects.}
    \label{fig: sim1_M1_spatial}%
\end{center}
\end{figure}

\begin{figure}[H]
\begin{center}
\includegraphics[scale=0.6]{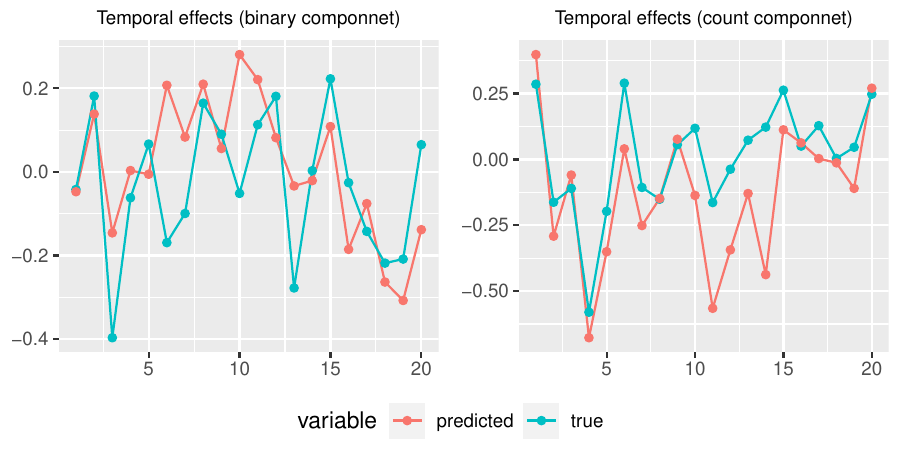}
    \caption{The true and fitted temporal effects in simulation 1 using the proposed model. }
    \label{fig: sim1_M1_temporal}%
\end{center}
\end{figure}

The results in Table \ref{res:simulation1} show that the 95\% credible intervals of the proposed model have covered the true values for all the parameters (i.e., $\alpha_0, \alpha_1, \beta_0,\beta_1$), while the intervals of the Bayesian ZINB model by \cite{neelon2019bayesian} do not cover the true values of $\alpha_0$ and $\beta_0$ and the intervals of the traditional ZINB do not cover the true values for parameters $\alpha_0$, $\alpha_1$ and $\beta_0$. Figure \ref{fig: sim1_M1_spatial} presents the map of the true and posterior mean spatial random effects in the binary component (in the top panel) and the count component (in the bottom panel). In both components, the fitted spatial pattern closely mirrors the true distribution, suggesting the proposed model recovers the underlying spatial pattern in the data. Figure~\ref{fig: sim1_M1_temporal} presents the true and fitted temporal effects for each temporal point. The predicted temporal patterns capture the trend of the true patterns in general. 

%%%%%%%%%%%%%%%%%%%%%%%%%%%%
%%%%% simulation 1 for S=200
%%%%%%%%%%%%%%%%%%%%%%%%%%%%

\begin{table}[H]
\footnotesize
\begin{tabular}{ p{1cm}|p{1cm}|p{3.5cm}|p{3.5cm}|p{3.5cm} }
 \hline
%  \multicolumn{4}{|c|}{City 4 (Testing dataset):V16-Method 2} \\
%  \hline
Para. & True Value & Proposed Model & Bayesian ZINB by
\citet{neelon2019bayesian} & Traditional ZINB\\
\hline
$\alpha_{0}$ &-0.25& -0.07 (-1.68, 1.57)& 0.25 (-0.05, 0.68)&0.62(0.43, 0.81) \\
$\alpha_{1}$ &0.25& 0.17 (0.06, 0.29)& 0.20 (0.07, 0.35)&-0.24(-0.32, -0.16) \\
$\beta_{0}$ &0.50& 0.82 (-0.72, 2.47)& 0.92 (0.69, 1.14)&-0.45(-0.91, 0.02) \\
$\beta_{1}$ &-0.25& -0.20 (-0.27, -0.12)& -0.20 (-0.28, -0.12)&-0.22(-0.37, -0.07) \\
% $\phi_{1s}$ &8.00& 1.07 (0.34, 2.48)&& \\
$l_{11}$ &0.35& 1.05 (0.64, 1.71)&& \\
$\sigma_{11}$ &0.50& 1.81 (1.40, 2.31)&& \\
$l_{12}$ &1.00& 3.45 (0.45, 4.92)&& \\
$\sigma_{12}$ &0.20& 0.77 (0.43, 1.19)&& \\
% $\phi_{2s}$  &8.00& 9.75 (0.48, 15.84)&& \\
$l_{21}$  &0.35& 0.32 (0.25, 1.44)&& \\
$\sigma_{21}$ &0.50& 1.21 (0.87, 2.15)&& \\
$l_{22}$ &1.00& 2.55 (0.15, 4.63)&& \\
$\sigma_{22}$ &0.20& 0.81 (0.42, 1.28)&& \\
$\sigma_{\epsilon_{11}}$ &0.05& 0.05 (0.03, 0.08)&& \\
$\sigma_{\epsilon_{12}}$ &0.05& 0.05 (0.03, 0.08)&& \\
$\sigma_{\epsilon_{21}}$ &0.05& 0.05 (0.03, 0.08)&& \\
$\sigma_{\epsilon_{22}}$ &0.05& 0.05 (0.03, 0.08)&& \\
$r$ &1.00&0.95 (0.63, 1.27)&0.73 (0.52, 1.14)& \\
\hline
\end{tabular}
% \captionof{table}
\caption{Parameter estimates and 95\% credible intervals for the proposed model, the spatiotemporal ZINB model by \cite{neelon2019bayesian}, and the traditional ZINB model in simulation 1. Note that the comparative models have different model assumptions and therefore do not have posterior results for hyperparameters used in the paper.}
\label{res:simulation1}
\end{table}

%%%%%%%%%%%%% M1 %%%%%%%%%

\subsubsection{Simulation 2: a large spatial dimension and one repetition in each sampling unit}
In simulation 2, we use the same setting as simulation 1 and increase the spatial dimension from $200$ to $500$, which results in a total sample size of $N=10,000$. The latent NNGP approximation is used to update $\sigma_{11}$ and $\sigma_{21}$ for the spatial random effects. The true and fitted results are shown in Table \ref{res:simulation2}. The 95\% credible intervals of the proposed model have covered the true values for all the parameters, while the intervals of the traditional ZINB model do not cover the true values for most parameters. We find that the Bayesian ZINB algorithm proposed by \citet{neelon2019bayesian} breaks in the first few iterations due to singularity issues which are caused by the high spatial dimension. 

\begin{figure}[H]
\begin{center}
\includegraphics[scale=0.6]{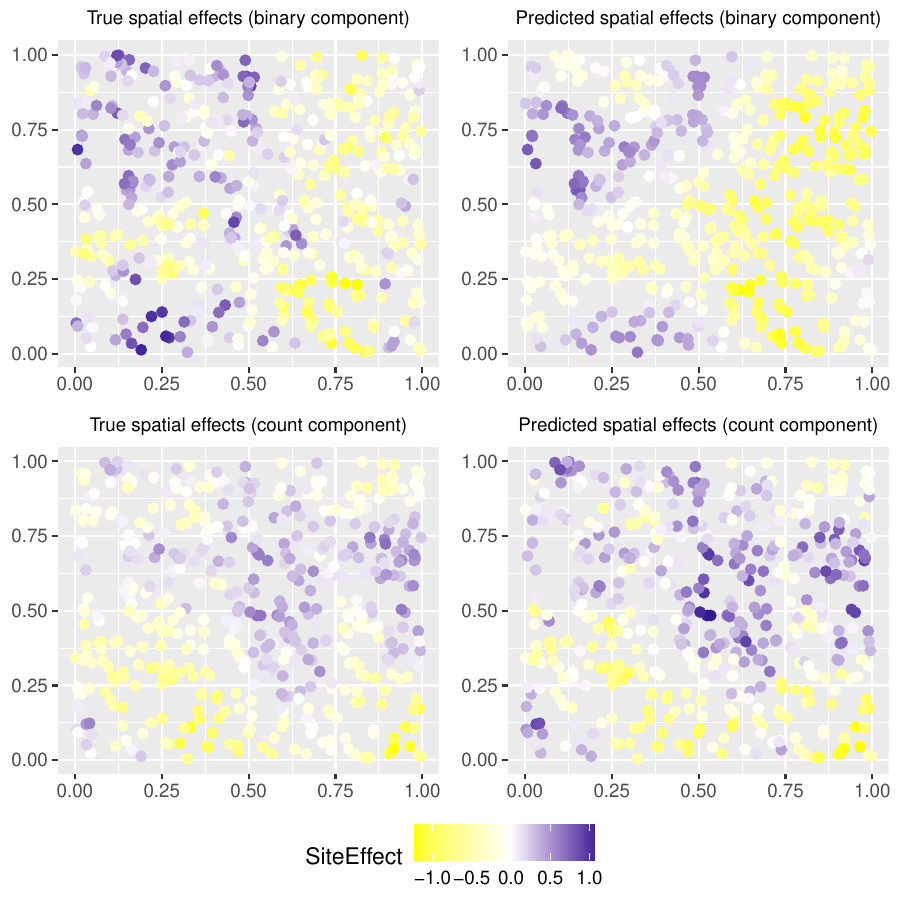}
    \caption{The true and fitted spatial effects in simulation 2 using the proposed model.}
    \label{fig: sim2_M1_spatial}%
\end{center}
\end{figure}

\begin{figure}[H]
\begin{center}
\includegraphics[scale=0.65]{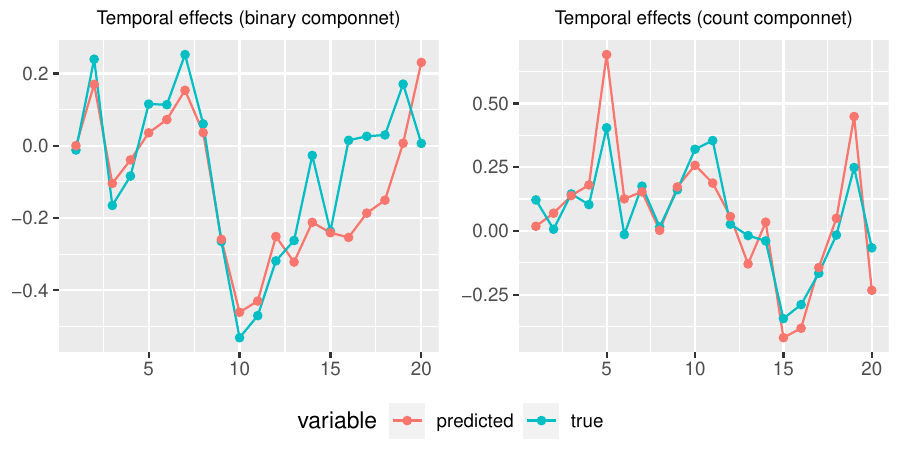}
    \caption{The true and fitted temporal effects in simulation 2 using the proposed model.}
    \label{fig: sim2_M1_temporal}%
\end{center}
\end{figure}

The true and posterior mean spatial and temporal random effects of simulation 2 are shown in Figure \ref{fig: sim2_M1_spatial} and \ref{fig: sim2_M1_temporal}. Both show that the fitted random effects are consistent with the true random effects.

%%%%%%%%%%%%%%%%%%%%%%%%%%%%
%%%%% simulation 2 for S=500
%%%%%%%%%%%%%%%%%%%%%%%%%%%%
\begin{table}[H]
\footnotesize
\begin{tabular}{ p{1cm}|p{2cm}|p{3.5cm}|p{3.5cm}}
 \hline
%  \multicolumn{4}{|c|}{City 4 (Testing dataset):V16-Method 2} \\
%  \hline
Para. & True Value & Proposed Model & Traditional ZINB\\
\hline
$\alpha_{0}$ &-0.25& 0.00 (-1.94, 1.84)&0.32(0.14, 0.50) \\
$\alpha_{1}$ &0.25& 0.20 (0.13, 0.28)&-0.23(-0.28, -0.17) \\
$\beta_{0}$ &0.50& 0.07 (-2.05, 2.14)&-0.20(-0.59, 0.20) \\
$\beta_{1}$ &-0.25& -0.23 (-0.28, -0.17)&-0.23(-0.34, -0.12) \\
% $\phi_{1s}$ &8.00& 0.11 (0.08, 0.22)& \\
$l_{11}$ &0.35& 3.02 (2.13, 3.54)& \\
$\sigma_{11}$ &0.50& 2.87 (2.34, 3.45)& \\
$l_{12}$ &1.00& 3.92 (2.37, 4.94)& \\
$\sigma_{12}$ &0.20& 0.77 (0.44, 1.22)& \\
% $\phi_{2s}$  &8.00& 6.70 (0.12, 14.77)& \\
$l_{21}$  &0.35& 0.39 (0.20, 2.89)& \\
$\sigma_{21}$ &0.50& 1.74 (1.22, 3.41)& \\
$l_{22}$ &1.00& 3.05 (1.16, 4.79)& \\
$\sigma_{22}$ &0.20& 0.83 (0.43, 1.33)& \\
$\sigma_{\epsilon_{11}}$ &0.05& 0.05 (0.03, 0.07)& \\
$\sigma_{\epsilon_{12}}$ &0.05& 0.05 (0.03, 0.08)& \\
$\sigma_{\epsilon_{21}}$ &0.05& 0.05 (0.03, 0.08)& \\
$\sigma_{\epsilon_{22}}$ &0.05& 0.05 (0.03, 0.08)& \\
$r$ &1.00& 1.08 (0.80, 1.29)& \\
\hline
\end{tabular}
% \captionof{table}
\caption{Parameter estimates and 95\% credible intervals for the proposed model and the traditional ZINB model in simulation 2.}
\label{res:simulation2}
\end{table}

%%%%%%%%%%%%% M1 %%%%%%%%%

\subsubsection{Simulation 3: dynamic number of repetitions in each sampling unit}
\begin{figure}[H]
\begin{center}
\includegraphics[scale=0.6]{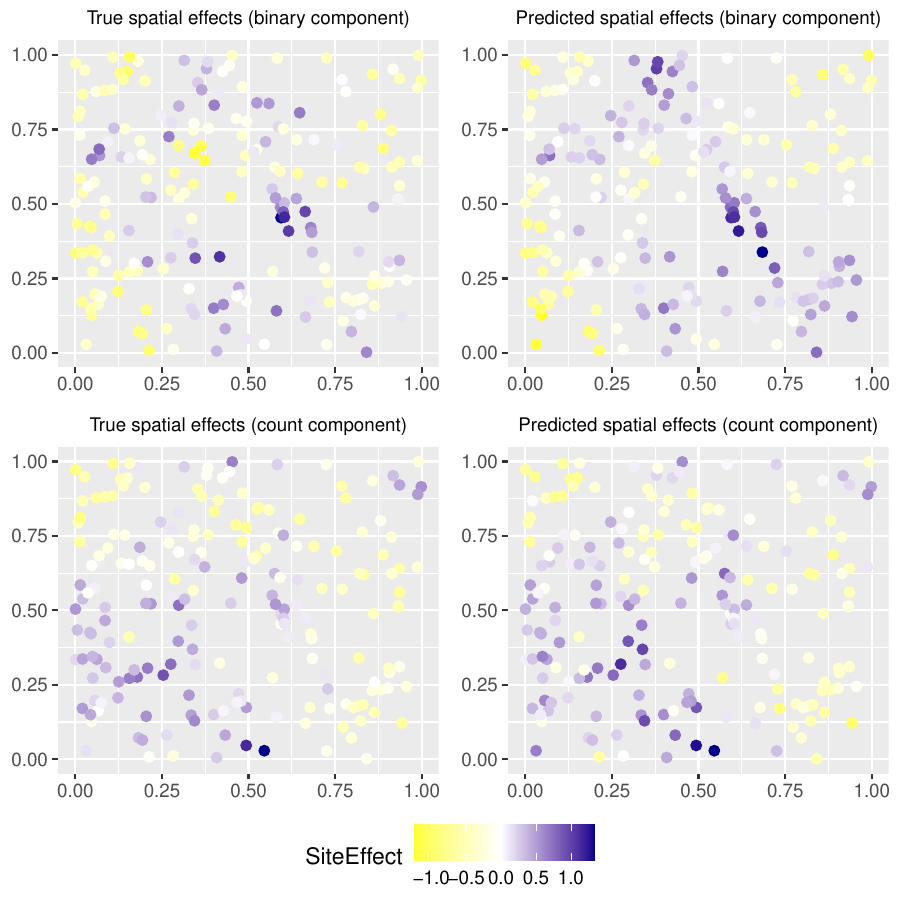}
    \caption{The true and fitted spatial effects in simulation 3 using the proposed model.}
    \label{fig: dynamic_M1_spatial}%
\end{center}
\end{figure}

\begin{figure}[H]
\begin{center}
\includegraphics[scale=0.6]{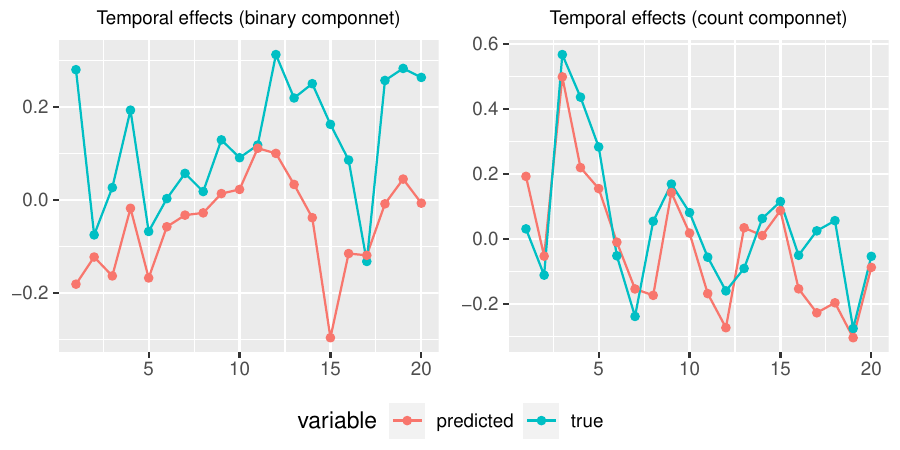}
    \caption{The true and fitted temporal effects in simulation 3 using the proposed model.}
    \label{fig: dynamic_M1_temporal}%
\end{center}
\end{figure}

In simulation 3, we run the simulation with a dynamic number of repetitions in each sampling unit. We set $S=200$ and $T=20$ like simulation~1. For each sampling unit $(s,t)$ the number of repetitions $n_{s,t}$ is randomly sampled from a Poisson distribution with mean $2$, which results in a total sample size of $N=7,867$. The posterior estimates and $95\%$ credible regions are presented in Table \ref{res:dynamic_M1}. 
% The proposed model outperforms both the Bayesian ZINB \citep{neelon2019bayesian} and the classical ZINB in the simulation settings 1 and 2. 
As the spatial size is relatively small and there are repetitions in each sampling unit in simulation 3, the Bayesian ZINB \citep{neelon2019bayesian} is comparable to the proposed model. Figure \ref{fig: dynamic_M1_spatial} and \ref{fig: dynamic_M1_temporal} show the true and fitted spatial and temporal random effects using the proposed model. The fitted random effects are better than those in simulation 1 and 2 where there is only one repetition per sampling unit.

\begin{table}[H]
\footnotesize
\begin{tabular}{ p{1cm}|p{1cm}|p{3cm}|p{3cm}|p{3cm}}
 \hline
Para. & True Value & Proposed Method & Bayesian ZINB by \citet{neelon2019bayesian} & Traditional ZINB\\
\hline
$\alpha_{0}$ &-0.25& -0.27 (-2.00, 1.44)& -0.34 (-0.48, -0.18)&0.45 (0.30, 0.60) \\
$\alpha_{1}$ &0.25& 0.18 (0.11, 0.26)& 0.19 (0.11, 0.26)&-0.20 (-0.26, -0.15) \\
$\beta_{0}$ &0.50& 0.31 (-1.68, 2.25)& 0.38 (0.2, 0.55)&-0.05 (-0.34, 0.25) \\
$\beta_{1}$ &-0.25& -0.20 (-0.26, -0.15)& -0.20 (-0.25, -0.15)&-0.21(-0.31, -0.12) \\
% $\phi_{1s}$ &8.00& 0.57 (0.29, 1.04)&& \\
$l_{11}$ &0.35& 1.32(0.98, 1.86)&& \\
$\sigma_{11}$ &0.50& 1.95 (1.53, 2.44)&& \\
$l_{12}$ &1.00& 3.91 (2.04, 4.96)&& \\
$\sigma_{12}$ &0.20& 0.74 (0.43, 1.14)&& \\
% $\phi_{2s}$  &8.00& 6.57 (0.32, 15.06)&& \\
$l_{21}$  &0.35& 0.39 (0.26, 1.77)&& \\
$\sigma_{21}$ &0.50& 1.37 (0.86, 2.59)&& \\
$l_{22}$ &1.00& 3.29 (0.99, 4.88)&& \\
$\sigma_{22}$ &0.20& 0.80 (0.47, 1.24)&& \\
$\sigma_{\epsilon_{11}}$ &0.05& 0.05 (0.03, 0.08)&& \\
$\sigma_{\epsilon_{12}}$ &0.05& 0.05 (0.03, 0.08)&& \\
$\sigma_{\epsilon_{21}}$ &0.05& 0.05 (0.03, 0.07)&& \\
$\sigma_{\epsilon_{22}}$ &0.05& 0.05 (0.03, 0.08)&& \\
$r$ &1.00& 1.19 (0.93, 1.46)& 1.13 (0.90, 0.55)& \\
% RMSPE & & & 12.32 & 3.56\\
% \hline
\hline
\end{tabular}
% \captionof{table}
\caption{Parameter estimates and 95\% credible intervals in simulation 3 using the proposed model and the spatiotemporal ZINB model.}
\label{res:dynamic_M1}
\end{table}

\subsection{COVID-19 fatality rates modeling}
COVID-19 has impacted populations around the world, with the fatality rate varying dramatically across counties of Florida \citep{karmakar2021association,khedhiri2021statistical,zhang2021association}.
Using statistical methods to identify factors which are associated with COVID-19 fatality/mortality rates is one of the most important topics. In the early stage of COVID-19, many counties were at low risk of having positive cases and did not have reported deaths. Hence, the proposed ZINB-NNGP model is a good option to model the mortality rate with excessive zeros. In this section, we apply the proposed model to understand the association of COVID-19 mortality at-risk and count rates with social vulnerability, adjusting other sociodemographic characteristics (i.e., health insurance coverage, urbanicity), population health care resources (i.e., primary care physicians), population health measures (i.e., life expectancy, obesity), and population density. We are not able to include the results of the Bayesian ZINB by \cite{neelon2019bayesian} for the COVID analysis because this Bayesian ZINB algorithm cracks after few iterations due to a singularity issue in high-dimensional spatial/temporal covariance matrix inversions, which also happened in simulation 2.

\begin{figure}[ht]
\begin{center}
\includegraphics[scale=.6]{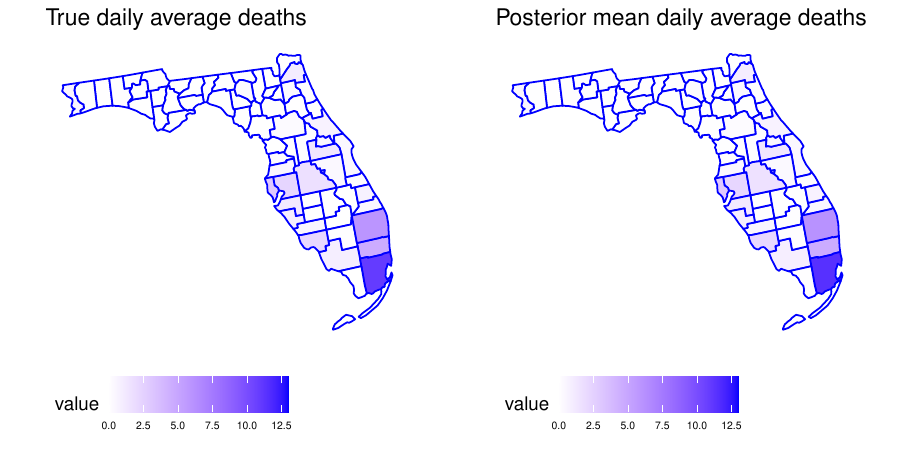}
    \caption{The true and fitted daily average deaths in Florida by counties.}
    \label{fig: COVID_SVI_y_pred_sp}%
\end{center}
\end{figure}

We consider the daily COVID-19 deaths in Florida at the county level reported from 3/25/2020 through 7/29/2020. More specifically, because of the variation in the schedule on which deaths are reported by days of week, we analyze the 7-day average deaths from the New York Times GitHub repository (\url{https://github.com/nytimes/covid-19-data}). 
The final analytic sample comprises 8,351 observations (67 counties across 127 study days).
The county level deaths in Florida during the selected time period are shown in the left plot of Figure \ref{fig: COVID_SVI_y_pred_sp}. 
There were approximately 74\% counties with  zero deaths reported in total, 14\% counties with an average of one death per day and 12\% counties with 2 ore more deaths per day.

Social vulnerability is measured by the social vulnerability index (SVI), which is a national and state-specific county ranking system and designed to assist public health officials in identifying communities in need of support and resources in an event of pandemic \citep{2018SVIFlorida}. The SVI has four different themes, namely, socioeconomic status, household composition, race/ethnicity/language, and housing/transportation, which comprise the overall SVI. We use the overall SVI in the analysis and categorize it to three levels: high ($>75$ percentile), moderate (between 25 percentile and 75 percentile), and low ($<25$ percentile). A higher level of overall SVI indicates more vulnerability, as suggested in the literature \citep{hughes2021county,bruckhaus2022covid}. We also adjust the following county-level characteristics in the model: other sociodemographic characteristics (i.e., health insurance coverage, urbanicity); population health care resources (i.e., primary care physicians); population health measures (i.e., life expectancy, obesity); and population density. The variables used in the analysis and the data sources are listed in Table \ref{table:data sources} \citep{dasgupta2020association}.

\begin{figure}[H]
    \centering
        \subfloat[\centering ]{\includegraphics[scale=0.7]{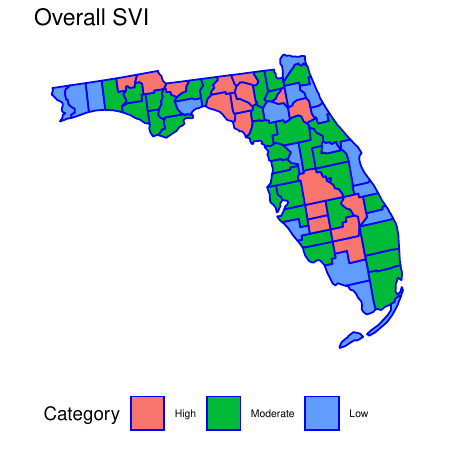}}
    %{\includegraphics[width=5cm]{Qing/network_Sep.png} }}%
    %\qquad
    \subfloat[\centering ]{{\includegraphics[scale=0.45]{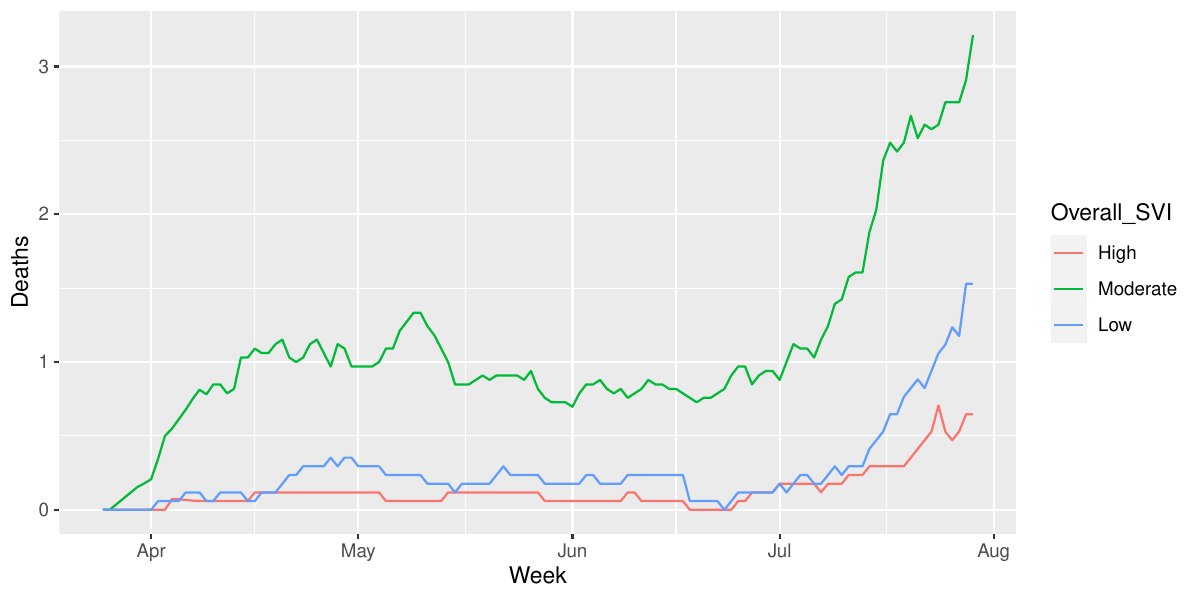}}}%

    \caption{\footnotesize  (a) The map of Florida with colors according to the county's overall SVI category. (b) The daily average deaths by overall SVI in Florida from 3/25/2020 to 7/29/2020.}%
    \label{fig: SVI map and true deaths}%
\end{figure}

\begin{figure}[H]
\begin{center}
\includegraphics[scale=0.5]{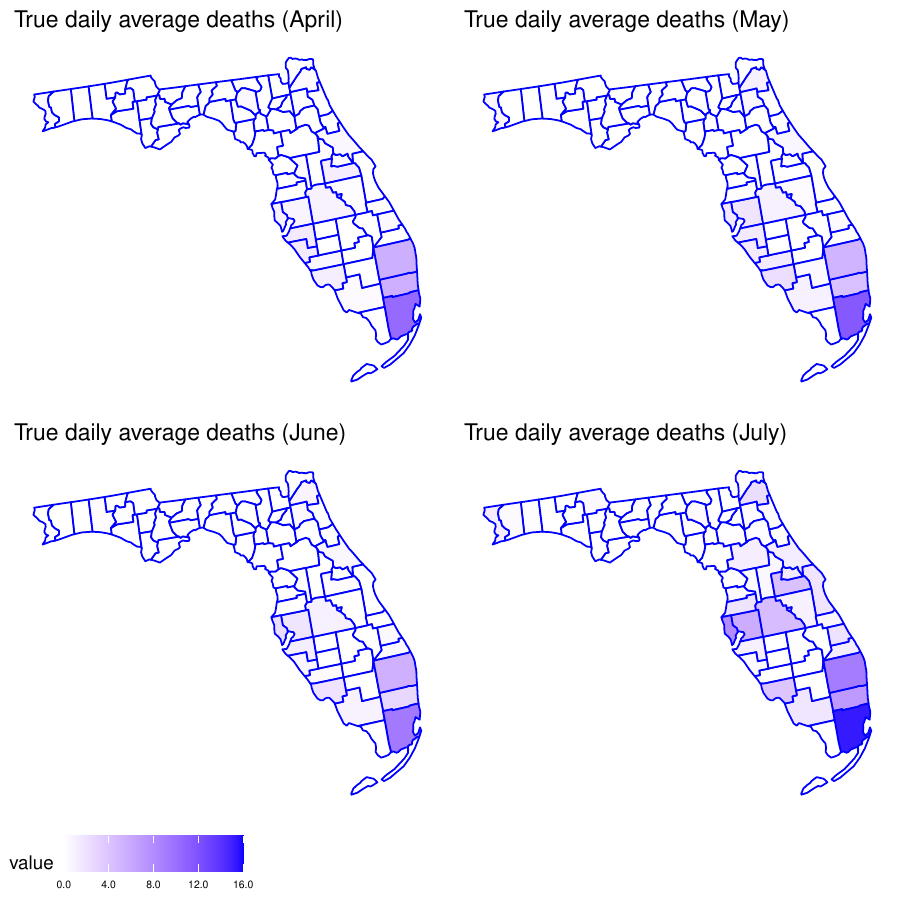}
    \caption{The true daily average deaths by month in the study period.}
    \label{fig: true deaths county map by month}%
\end{center}
\end{figure}

A map of the overall SVI among Florida counties (panel (a)) and a time series plot of the reported daily average deaths for each SVI category (panel (b)) are shown in Figure \ref{fig: SVI map and true deaths}. In panel (a), it is shown that most counties have moderate overall SVI (green color in the map). In panel (b), the counties with moderate overall SVI have more deaths than the high and low group. It is different from the finding that counties with high SVI had more COVID-19 deaths in other research focused on the U.S. or other locations \citep{karmakar2021association, neelon2022multivariate}. To further understand how the COVID-19 deaths spread across the three SVI categories, we show the map of the daily average deaths in four different months in Figure \ref{fig: true deaths county map by month}. Counties at a higher risk such as Miami-Dade and Hillsborough have more COVID-19 deaths and they have moderate overall SVI scores.

%%%%%%%%%%%%% Variables and data sources
\begin{table}[H]

\begin{tabular}{ p{4.4cm}|p{1.6cm}|p{6.5cm}}
 \hline
%  \multicolumn{4}{|c|}{City 4 (Testing dataset):V16-Method 2} \\
Variable&Year&Source\\\hline\hline
COVID-19 deaths&2020&The New York Times\\\hline
Social Vulnerability Index and component measures&&\\\hline
Overall Social Vulnerability Index&2014-2018&Centers for Disease Control and Prevention\\
Socioeconomic status index&2014-2018&Centers for Disease Control and Prevention\\
Poverty rate&2014-2018&Centers for Disease Control and Prevention\\
% Unemployment rate&2014-2018&Centers for Disease Control and Prevention\\
% Income&2014-2018&Centers for Disease Control and Prevention\\
% Educational attainment&2014-2018&Centers for Disease Control and Prevention\\
Household characteristics and disability index&2014-2018&Centers for Disease Control and Prevention\\
65 years or older&2014-2018&Centers for Disease Control and Prevention\\
% 17 years or younger&2014-2018&Centers for Disease Control and Prevention\\
% Disability&2014-2018&Centers for Disease Control and Prevention\\
% Single parent household&2014-2018&Centers for Disease Control and Prevention\\
Minority status and language index&2014-2018&Centers for Disease Control and Prevention\\
Minority&2014-2018&Centers for Disease Control and Prevention\\
% Limited English proficiency&2014-2018&Centers for Disease Control and Prevention\\
Housing type and transportation index&2014-2018&Centers for Disease Control and Prevention\\
\hline
No health insurance coverage&2015-2019&US Census Bureau\\
Public transportation to commute to work&2015-2019&US Census Bureau\\
Urbanicity&2013&US Department of Agriculture\\
Total primary care physicians&2019&Health Resources \& Services Administration\\
Life expectancy&2016-2018&Robert Wood Johnson Foundation\\
Obesity&2016&Robert Wood Johnson Foundation\\
Population Density&2014-2018&Centers for Disease Control and Prevention\\
% Geographic area&2018&Centers for Disease Control and Prevention\\
% PCR tests rate&2021&The COVID Tracking Project\\
\hline
\end{tabular}
% \captionof{table}
\caption{Data sources.}
\label{table:data sources}
\end{table}

Table \ref{table: overall SVI results_zinb} shows the estimated coefficients as well as the 95\% credible intervals in the binary and count component. Counties with higher overall SVI levels are more likely to report deaths (binary component). This is consistent with the previous research \citep{karmakar2021association}. Meanwhile, the percentage of people under age 65 without health insurance is associated with a higher risk of reporting deaths, which is consistent with finding from other studies \citep{mountantonakis2021association,dupre2021county}. It is also found that obesity is associated with a lower risk of reporting deaths, although studies have suggested that obesity is a risk factor for death from COVID-19 \citep{tartof2020obesity, rottoli2020important}. This may be due to potentially confounding effects of obesity with the overall SVI \citep{an2015social}. Population density is associated with a higher risk of reporting deaths and more deaths, which is consistent with other research \citep{neelon2021spatial,foo2021global}.

\begin{table}[H]
\footnotesize
\begin{tabular}{ p{4.5cm}|p{4cm}|p{4cm} }
 \hline
Variable&Binary component&Count component\\
\hline
Overall SVI& 1.06 (0.011, 2.109)& 0.418 (-0.412, 1.229)\\
No health insurance coverage& 0.308 (0.114, 0.492)& 0.014 (-0.118, 0.148)\\
Urbanicity& 0.021 (-0.471, 0.484)& -0.261 (-0.706, 0.153)\\
Total primary care physicians& -0.017 (-0.038, 0.004)& -0.005 (-0.019, 0.009)\\
Life expectancy& -0.049 (-0.14, 0.051)& -0.009 (-0.181, 0.154)\\
Obesity& -0.248 (-0.399, -0.104)& -0.062 (-0.187, 0.062)\\
Population density& 1.387 (0.658, 2.109)& 1.415 (0.883, 1.993)\\
\hline
\end{tabular}
% \captionof{table}
\caption{The parameter estimates and 95\% credible intervals of the Florida counties' sociodemographic and health characteristics with COVID-19 deaths.}
\label{table: overall SVI results_zinb}
\end{table}

\begin{figure}[H]
    \centering
      \subfloat[\centering ]{\includegraphics[scale=0.37]{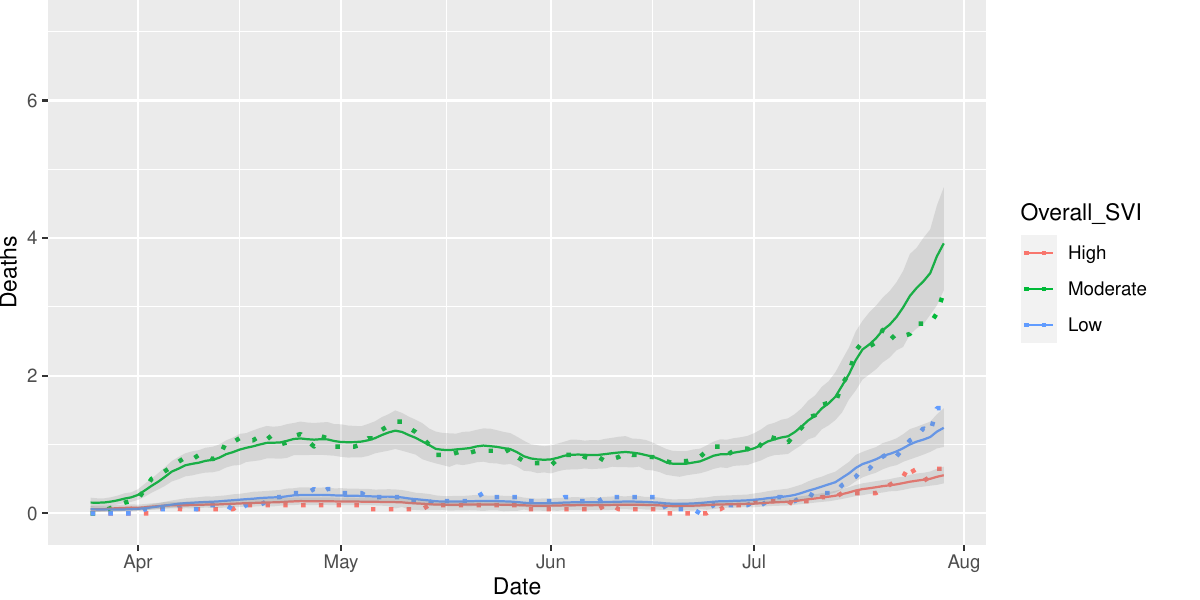}}
    %{\includegraphics[width=5cm]{Qing/network_Sep.png} }}%
    %\qquad
    \subfloat[\centering ]{{\includegraphics[scale=0.37]{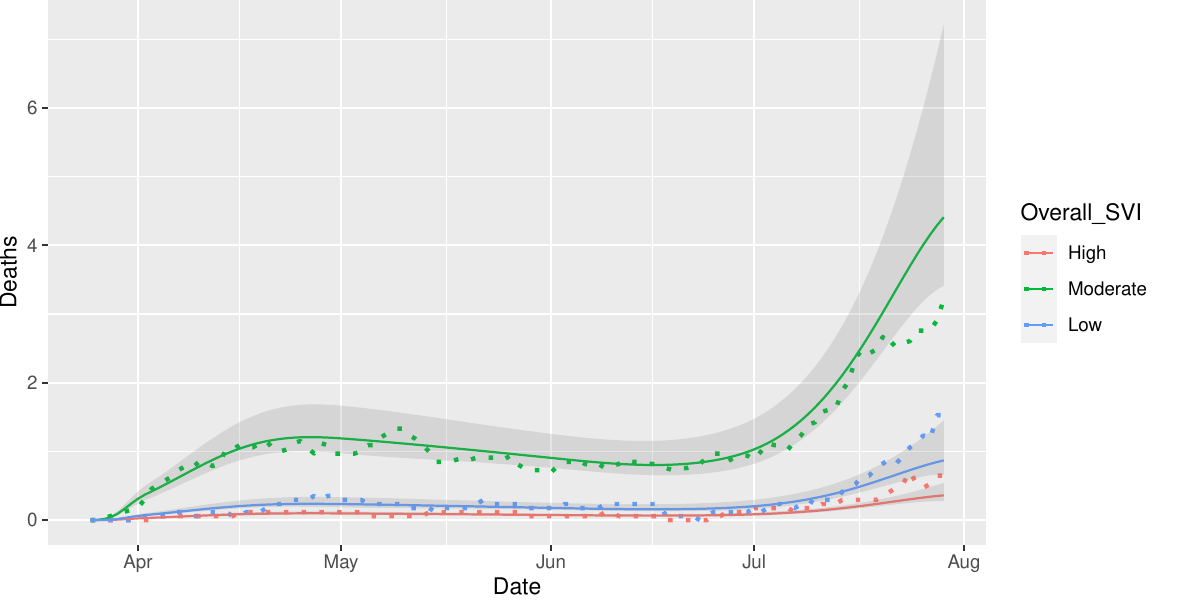}}}%
    \caption{\footnotesize  The true and fitted daily average deaths in Florida by SVI category using the proposed model in panel (a) and the B-splines spatiotemporal model proposed by \cite{neelon2022multivariate} in panel (b). The dotted lines are the true daily average, solid lines are the fitted daily average, and the grey ribbons are the 95\% credible region.}%
    \label{fig: y_pred by SVI}%
\end{figure}

The fitted daily average death by overall SVI category is shown in panel~(a) of Figure \ref{fig: y_pred by SVI}. The fitted trend (solid line) is consistent with the actual (dotted line). In the earlier time (March 2020 through the middle of July 2020), there is approximately one death each day for counties with moderate SVI and zero deaths for counties with low or high SVI. Starting from July 2020, the death counts has increased for all counties in Florida, and counties in the moderate category had the most significant increase than others. We also apply a recently developed spatiotemporal model (not in a zero-inflated setting) \citep{neelon2022multivariate} which used B-splines to model temporal effects for the COVID-19 analysis. We notice that it could be difficult to set the location and number of knots appropriately. Its fitted daily average deaths by SVI category are shown in panel (b) of Figure \ref{fig: y_pred by SVI}. There is a higher predictive uncertainty at the end of the temporal period for all three overall SVI categories.

\begin{figure}[H]
\begin{center}
\includegraphics[scale=0.5]{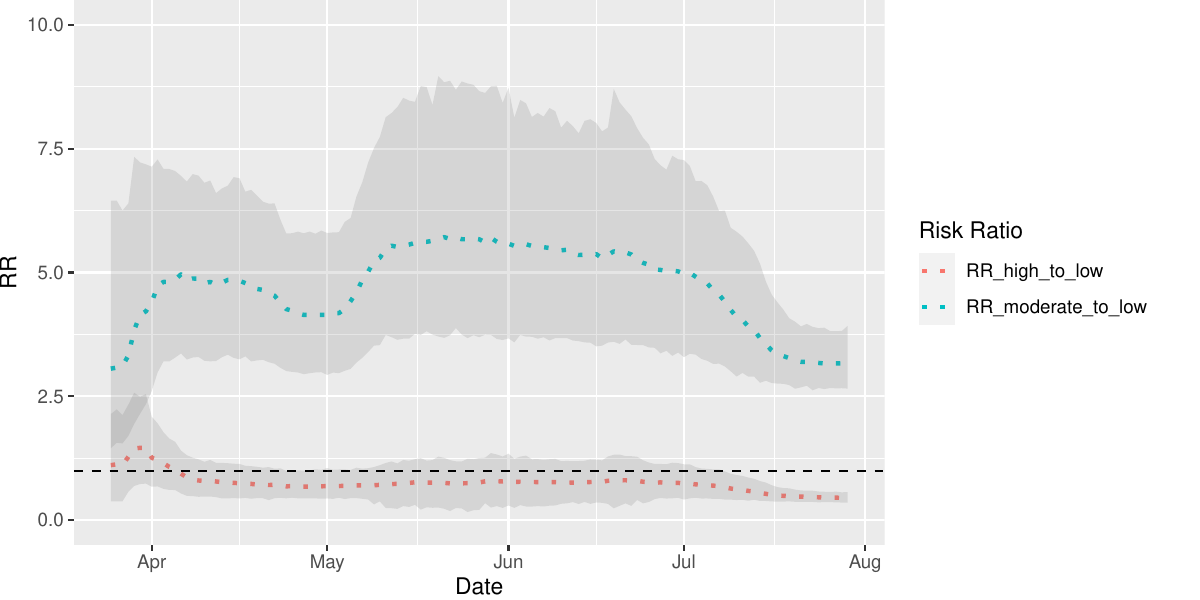}
    \caption{The RRs for counties with high and moderate social vulnerability versus low social vulnerability. The dotted lines are posterior mean RRs and the grey ribbons were the 95\% credible regions. RRs $>$ 1 indicate higher risk. The black dashed line is the reference line with risk ratio $=1$.}
    \label{fig: risk ratio}%
\end{center}
\end{figure}

To further compare the death counts across the three SVI categories, risk ratios (RRs) for the high and moderate SVI versus the low SVI are shown in Figure \ref{fig: risk ratio}, the counties in moderate SVI category have a higher risk of reporting deaths than the counties in the low SVI category (the blue dotted line). The counties in the high SVI category have a similar risk of reporting death with counties in the low SVI category between March 2020 and June 2020 and lower risk in July 2020 (red dotted line).
The county level observed average death counts and posterior mean average death counts in the overall SVI model is shown in Figure \ref{fig: COVID_SVI_y_pred_sp}. The proposed model identifies counties with high death counts such as Miami-Dade.
\begin{figure}[H]
    \centering
        \subfloat[\centering ]{\includegraphics[scale=0.3]{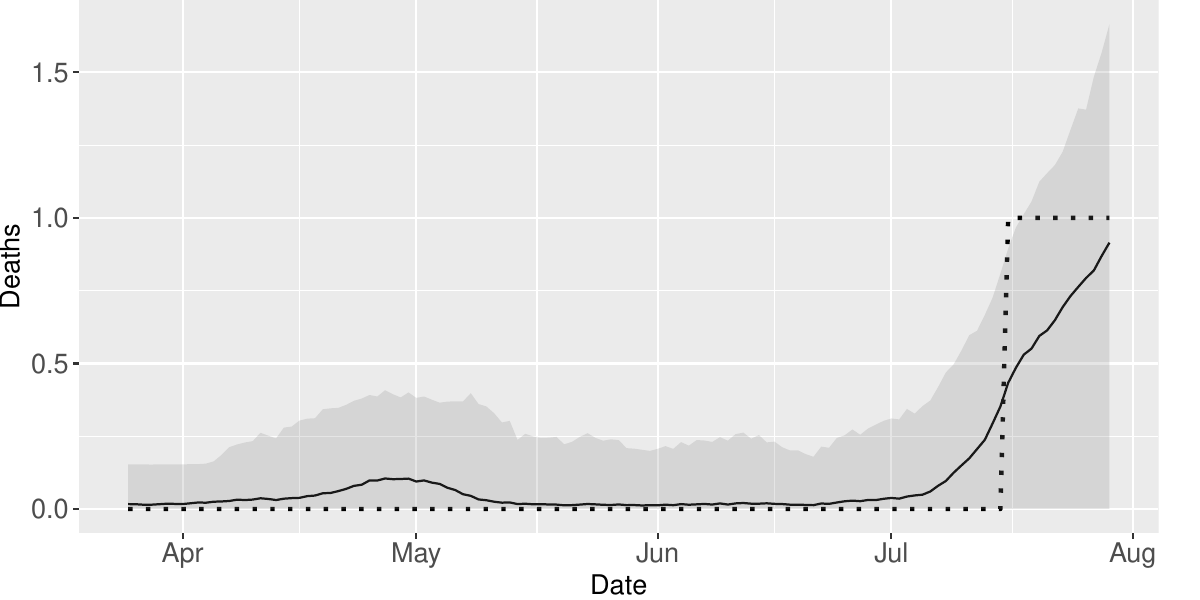}}
    %{\includegraphics[width=5cm]{Qing/network_Sep.png} }}%
    \qquad
    \subfloat[\centering ]{{\includegraphics[scale=0.3]{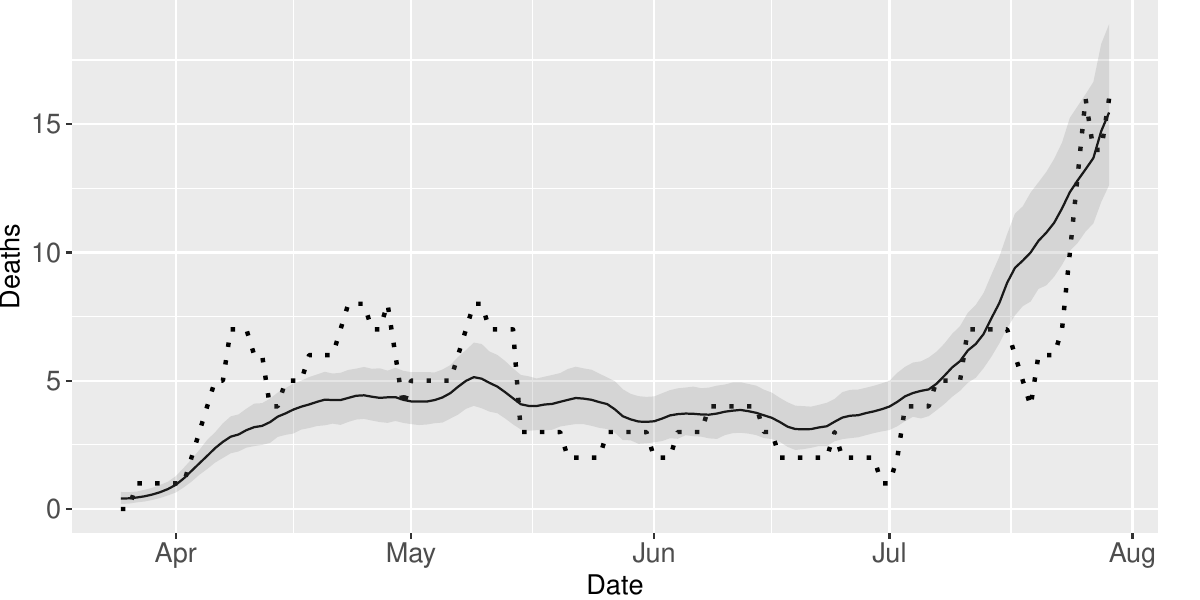}}}%
    \qquad
    \subfloat[\centering ]{{\includegraphics[scale=0.3]{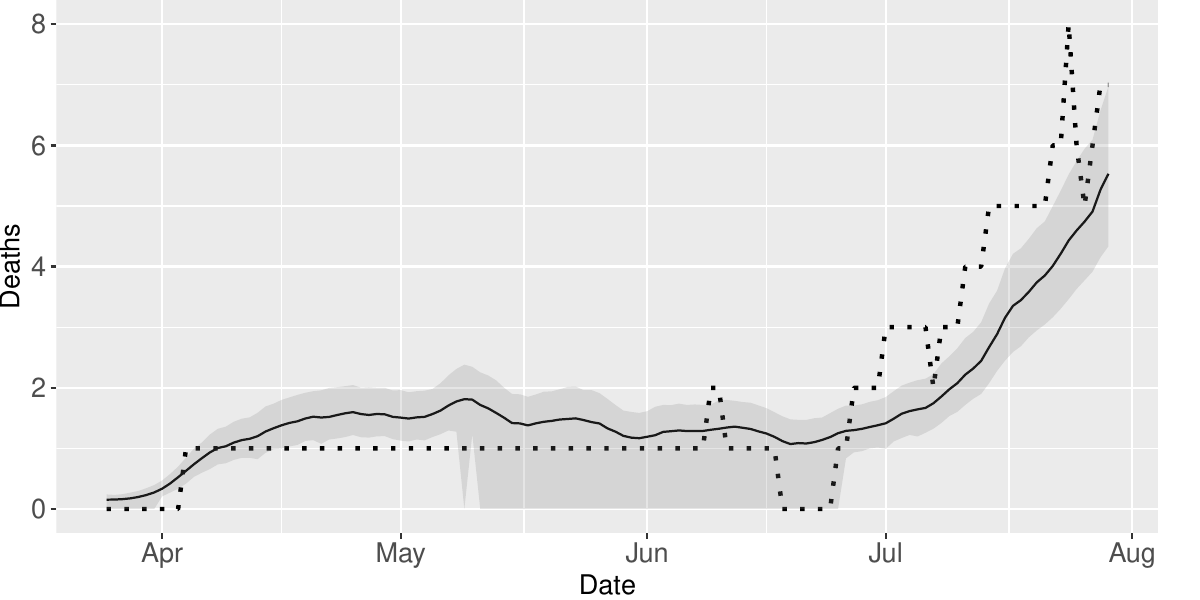}}}%
    \qquad
    \subfloat[\centering ]{{\includegraphics[scale=0.3]{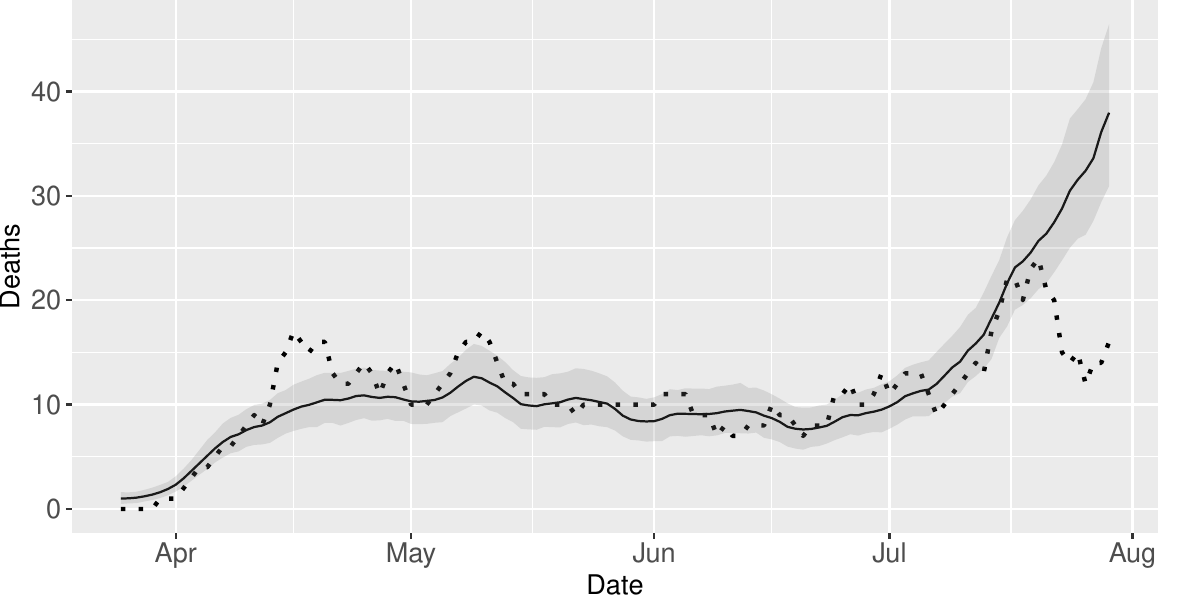}}}%
    \caption{\footnotesize  The true and posterior mean daily deaths in Hernando county (a), Broward county (b), Polk county (c), and Miami-Dade county (d) from 3/25/2020 to 7/29/2020 using our proposed method. The dotted lines are the true deaths and the solid lines are the fitted deaths. The grey ribbons are the 95\% credible intervals.}%
    \label{fig: county examples}%
\end{figure}
\begin{figure}[H]
    \centering
        \subfloat[\centering ]{\includegraphics[scale=0.3]{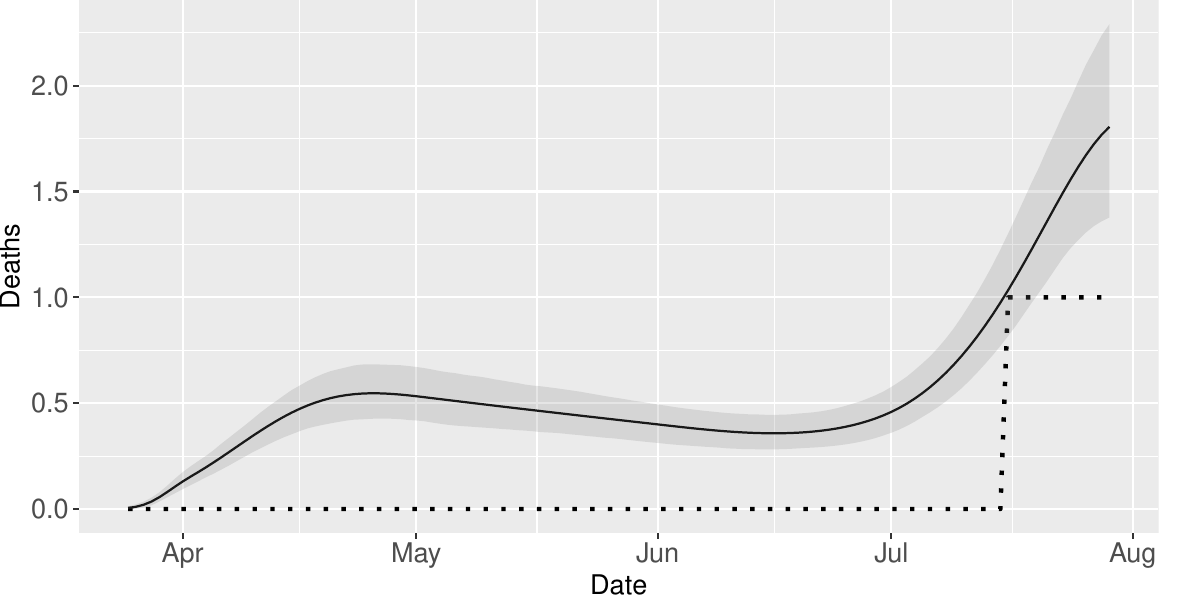}}
    %{\includegraphics[width=5cm]{Qing/network_Sep.png} }}%
    \qquad
    \subfloat[\centering ]{{\includegraphics[scale=0.3]{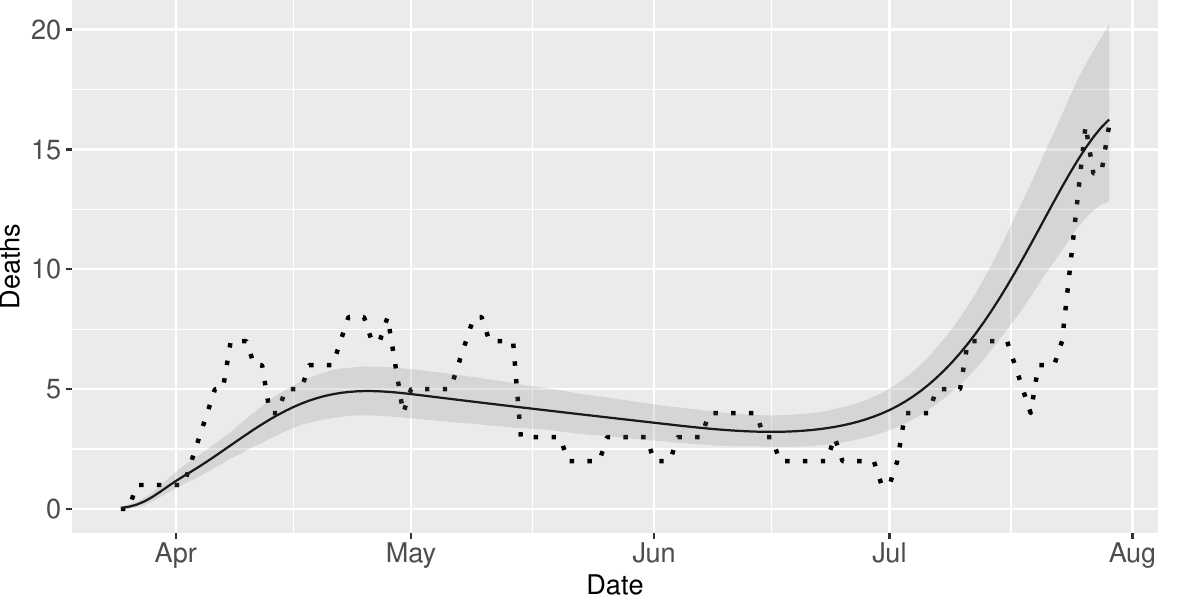}}}%
    \qquad
    \subfloat[\centering ]{{\includegraphics[scale=0.3]{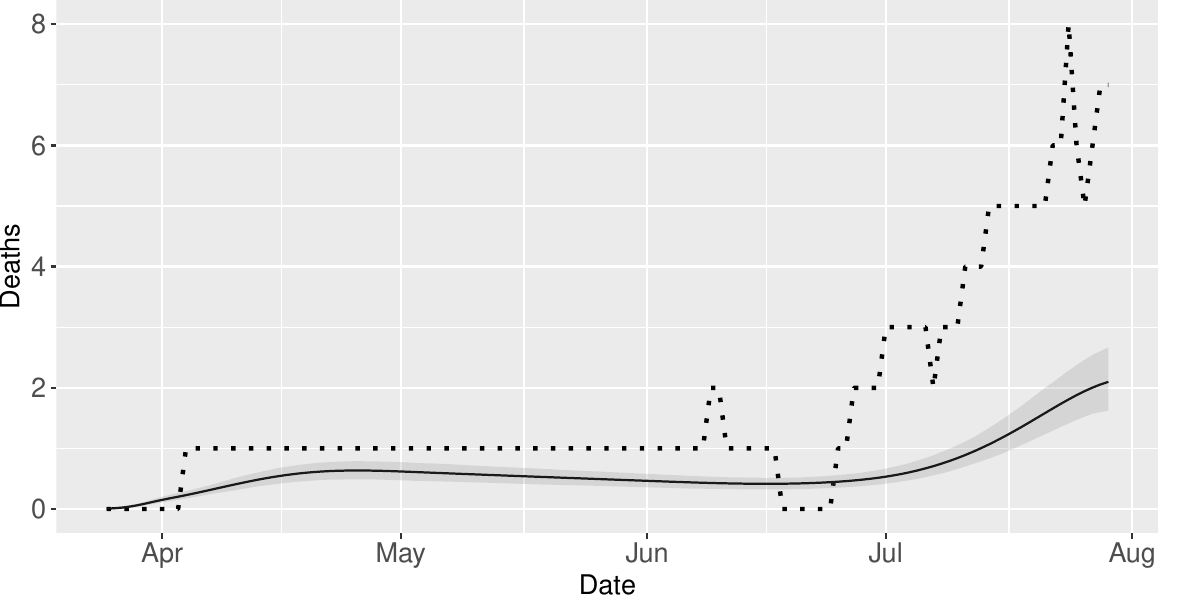}}}%
    \qquad
    \subfloat[\centering ]{{\includegraphics[scale=0.3]{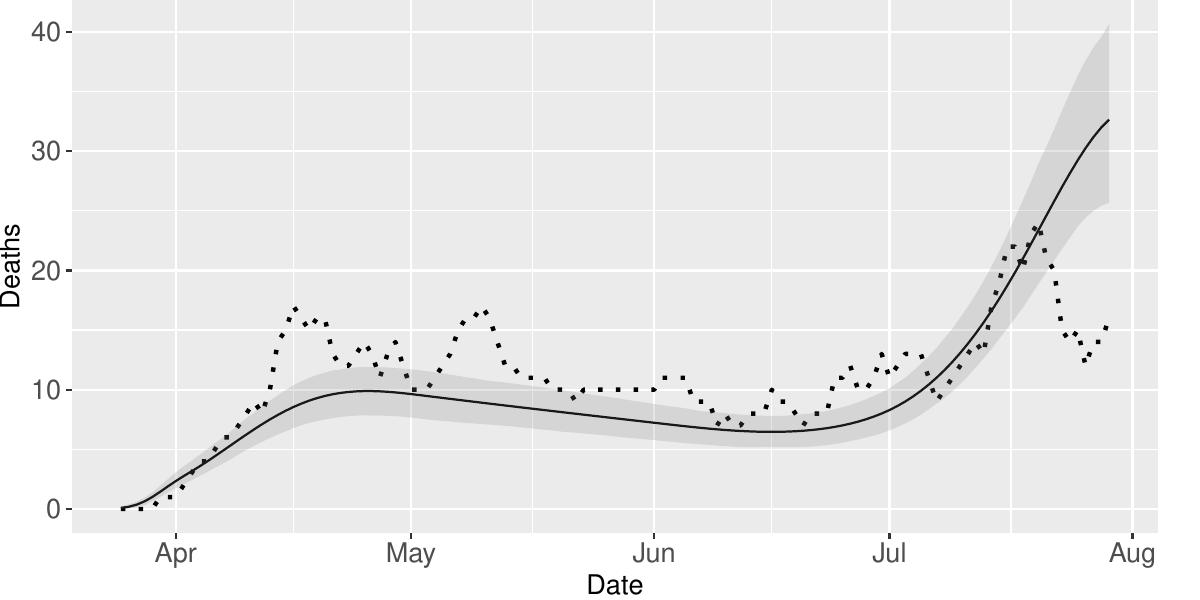}}}%
    \caption{\footnotesize  The true and posterior mean daily deaths in Hernando county (a), Broward county (b), Polk county (c), and Miami-Dade county (d) from 3/25/2020 to 7/29/2020 using the B-splines spatiotemporal model proposed by \cite{neelon2022multivariate}. The dotted lines are the true deaths and the solid lines are the fitted deaths. The grey ribbons are the 95\% credible intervals.}%
    \label{fig: Bspline county examples}%
\end{figure}

In Figure \ref{fig: county examples}, we select four counties from the three SVI categories to show the individual fitted daily deaths: Hernando county (moderate SVI), Broward county (high SVI), Polk county (low SVI), and Miami-Dade county (moderate SVI). The fitted curves capture the overall patterns but are generally more smooth than the observed death curves which are more wiggly. The model does not capture the downward trend in Miami-Dade (see panel~(d) of Figure~\ref{fig: county examples}) in the later period of July 2020. This may be related to emergency orders of closing certain indoor spaces and outdoor spaces where groups of people congregated without physical distancing in July 2020, which are not included in the analysis.
For comparisons, we show the fitting results using the spatiotemporal model proposed in \citet{neelon2022multivariate} in Figure \ref{fig: Bspline county examples}. The posterior mean daily deaths do not capture the trends very well and the 95\% CIs fail to cover most the true values for Hernando county in panel (a), Polk county in panel (c), and Miami-Dade county in panel (d). The result for Broward county in panel (b) is similar to that of the proposed model.

\section{Discussion}
\label{sec4}
We have introduced a Bayesian framework for zero-inflated negative binomial regression models with spatiotemporal effects. The proposed model is able to simultaneously incorporate a flexible structure for the spatial effects and temporal for the binary and count component.
The Gibbs sampler is efficient for posterior inference via the P\'{o}lya-Gamma data augmentation and latent NNGP conditioned on these latent variables through the Bayesian hierarchical inference. The use of the latent NNGP to approximate the covariance inverse matrix empowers the proposed to be feasible when the spatial or temporal dimension is large. 
Our simulations suggest that the proposed model is comparable to existing methods when such comparisons are available and more applicable in a higher dimension. 
It also outperforms other methods when there is only one repetition in each sampling unit. 

There are several potential improvements for future work. There is potential for selecting important variables in this Bayesian framework.
For example, by using multivariate Bayesian models with polynomial-tailed shrinkage priors \citep{wang2023mixedtype} for the covariance matrix, one can derive variable selection methods. Such methods would be useful to model ultrahigh dimensional spatiotemporal data. Additionally, the proposed model assumes a common temporal effect across different spatial locations and a common spatial effect across different temporal points. Future study could be focused on dynamic settings that capture the interaction between spatial and temporal effects.

\section{Software}
\label{sec5}

Software in the form of R code, together with a sample
input data set and complete documentation is available on
request from the corresponding author (hsin.huang@ucf.edu).

\section*{Acknowledgments}

The authors thank the reviewers for providing valuable comments. We would like to acknowledge support for this project
from the National Science Foundation (NSF grant DMS-1924792).
{\it Conflict of Interest}: None declared.

\section*{Appendix}
\subsection*{The step-by-step posterior sampling procedure}\label{section: bayesian inference}
The update algorithm for parameters of interests as well as random effects and hyperparameters are shown below. In each component, we update parameters and random effects together as the posterior distribution are both Gaussian, which can facilitate convergence of the MCMC algorithm in simulation studies.\\

Step 1: Update the latent at-risk indicators\\
The update for $W_j$ depends on the value of $y_j$.
If $y_j > 0$, then the $j$th subject belongs to the at-risk class and $W_j = 1$. If $y_j = 0$, $W_j$ is updated by drawing from a Bernoulli distribution with probability
    % then we observe either a structural zero (implying that $W_j = 0$) or an at-risk zero (implying $W_j = 1$). Conversely,  Based on Bayes’ Theorem, the at-risk probability (i.e., $W_j=1$) conditional on $y_j>0$ is as follows 
% $W_j$ is updated by drawing from a Bernouli distribution with probability
\begin{align*}
    Pr(W_j=1\mid y_j=0,\text{rest}) = \frac{\pi_j \nu_j^r}{1-\pi_j(1- \nu_j^r)},
\end{align*}
where $\pi_j = \frac{\exp(\eta_{1j})}{1+\exp(\eta_{1j})}$ is the unconditional probability that $W_j=1$ and $\nu_j=1-\psi_j$.\\

Step 2: Update $\alpha, \vec{a}, \vec{b}$\\
Denote $[\alpha, \vec{a}, \vec{b}]^T$ by $\phi_1$. Assume a $\mathcal{N}(\phi_0,\Sigma_0)$ prior, we update $\phi_1$ from the posterior distribution in two steps:
$ \omega_{1j}\mid\eta_{1j} \sim \mathcal{PG}(1,\eta_{1j})$ and $
    \phi_1 \mid \Omega_1, z_1 \sim \mathcal{N}(\mu, \Sigma),$
where 
\begin{eqnarray*}
  \Sigma &=& (\Sigma_0^{-1} +V^{T}
\Omega_1 V)^{-1},\\
\mu &=& \Sigma
\left(\Sigma_0^{-1}\phi_0 +V^{T}
\Omega_1 (z_1-V_1 \vec{\epsilon}_{11} - V_2 \vec{\epsilon}_{12})\right),
\end{eqnarray*}
and $z_{1j} =\frac{W_j -1/2}{\omega_{1j}}$ is the latent variable, $ \Omega_1 = \text{diag}(\omega_{1j})$, and $V=[X, V_1, V_2]$ is the augmented design matrix.\\

Step 3: Update $\vec{\epsilon}_{11}$\\
Conditional on $\alpha, \vec{a}, \vec{b},\vec{\epsilon}_{12}$, assuming a $\mathcal{N}(0,\sigma^2_{\epsilon_{11}})$ prior, update $\vec{\epsilon}_{11}$ from a $\mathcal{N}(\mu,\Sigma)$
distribution where
\begin{eqnarray*}
  \Sigma &=&
\left(\frac{1}{\sigma^2_{\epsilon_{11}}}I +V_1^{T}
\Omega_1 V_1\right)^{-1},\\
\mu &=& \Sigma
\left(V_1^{T}
\Omega_1(z_1 -X \alpha - V_1 \vec{a} -V_2 \vec{b} - V_2 \vec{\epsilon}_{12})\right).  
\label{pos_beta}
\end{eqnarray*}\\

Step 4: Update $\vec{\epsilon}_{12}$\\
Conditional on $\alpha, \vec{a}, \vec{b},\vec{\epsilon}_{11}$, assuming a $\mathcal{N}(0,\sigma^2_{\epsilon_{12}})$ prior, update $\vec{\epsilon}_{12}$ from a $\mathcal{N}(\mu,\Sigma)$
distribution, where
\begin{eqnarray*}
  \Sigma &=&
\left(\frac{1}{\sigma^2_{\epsilon_{12}}}I +V_2^{T}
\Omega_1 V_2\right)^{-1},\\
\mu &=& \Sigma
\left(V_2^{T}
\Omega_1(z_1 -X \alpha - V_1 \vec{a} -V_2 \vec{b} - V_1 \vec{\epsilon}_{11})\right).  
\label{pos_beta}
\end{eqnarray*}\\

Step 5: Update $l_{11}, \sigma_{11}$\\
First, we use the Metropolis-Hasting algorithm to update $l_{11}$. Using a normal proposal distribution which is symmetric, the acceptance probability can be written as follows:
\begin{align*}
    A & = \min \left( 1, \frac{\pi(l^*_{11})}{\pi(l_{11})} \right)\\
      & = \min \left( 1, \frac{p(\vec{a}\mid K_{\sigma_{11},l^*_{11}} (h_1, \ldots, h_S))p(l^*_{11}\mid a_{l_1} , b_{l_1})}{p(\vec{a}\mid K_{\sigma_{11},l_{11}} (h_1, \ldots, h_S))p(l_{11}\mid a_{l_1} , b_{l_1})} \right).
\end{align*}
Then, $\sigma_{11}$ can be sampled from the full conditional as
$$
\sigma^2_{11} \sim \mathcal{IG}\left(a_{\sigma_1} + \frac{S}{2}, b_{\sigma_1} + \frac{\vec{a}^T (K_{\sigma_{11},l_{11}} (h_1, \ldots, h_S))^{-1} \vec{a}}{2}\right)
$$\\
The posterior distribution of $\sigma_{11}$ requires the inverse of 
$K_{\sigma_{11},l_{11}} (h_1, \ldots, h_S)$, which may be a computation bottleneck when the size of spatial locations is large. Hence, we apply the latent conjugate NNGP to approximate the inverse matrix.\\

Step 6: Update $l_{12}, \sigma^2_{12}$\\
The update for $l_{12}, \sigma^2_{12}$ is similar to the one for $l_{11}, \sigma^2_{11}$. Depending on the size of temporal points, the latent conjugate NNGP can also be applied in the update for $\sigma^2_{12}$ if $T$ is large. First, we use the Metropolis-Hasting algorithm to update $l_{12}$. Using a normal proposal distribution which is symmetric, the acceptance probability can be written as follows:
\begin{align*}
    A & = \min \left( 1, \frac{\pi(l^*_{12})}{\pi(l_{12})} \right)\\
      & = \min \left( 1, \frac{p(\vec{b}\mid K_{\sigma_{12},l^*_{12}} (w_1, \ldots, w_T))p(l^*_{12}\mid a_{l_2} , b_{l_2})}{p(\vec{b}\mid K_{\sigma_{12},l_{12}} (w_1, \ldots, w_T))p(l_{12}\mid b_{l_2} , b_{l_2})} \right).
\end{align*}
Then, $\sigma_{12}$ can be sampled from the full conditional as
$$
\sigma^2_{12} \sim \mathcal{IG}\left(a_{\sigma_1} + \frac{T}{2}, b_{\sigma_2} + \frac{\vec{b}^T (K_{\sigma_{12},l_{12}} (w_1, \ldots, w_T))^{-1} \vec{b}}{2}\right).
$$\\

Step 7: Update $\sigma_{\epsilon_{11}}$\\
Assuming a $\mathcal{IG}(a_{\epsilon},b_{\epsilon})$ prior, draw $\sigma_{\epsilon_{11}}$ from the posterior distribution:
\begin{align*}
    \sigma^2_{\epsilon_{11}} \sim \mathcal{IG}\left(a_{\epsilon} + \frac{S}{2}, b_{ \epsilon} + \frac{\sum_{s=1}^{S}\epsilon^2_{11s}}{2}\right).
\end{align*}\\

Step 8: Update $\sigma_{\epsilon_{12}}$\\
Assuming a $\mathcal{IG}(a_{\epsilon},b_{\epsilon})$ prior, draw $\sigma_{\epsilon_{12}}$ from the posterior distribution:
\begin{align*}
    \sigma^2_{\epsilon_{12}} \sim \mathcal{IG}\left(a_{\epsilon} + \frac{T}{2}, b_{ \epsilon} + \frac{\sum_{t=1}^{T}\epsilon^2_{12t}}{2}\right).
\end{align*}\\

Step 9: Update $\beta, \vec{c}, \vec{d}$\\
Denote $[\beta, \vec{c}, \vec{d}]^T$ by $\phi_2$. Assume a $\mathcal{N}(\phi_0,\Sigma_0)$ prior, we update $\phi_2$ in two steps:
\begin{eqnarray*}
    \omega_{2j}|\eta_{2j} &\sim& \mathcal{PG}(y_j+r,\eta_{2j}), \forall j \text{ s.t. } W_j=1\\
    \phi_2 | \Omega_2, z_2 &\sim& \mathcal{N}(\mu, \Sigma),
\end{eqnarray*} 
where 
\begin{eqnarray*}
  \Sigma &=& (\Sigma_0^{-1} +V^{*T}
\Omega_2 V^*)^{-1}\\
\mu &=& \Sigma
\left(\Sigma_0^{-1}\phi_0 +V^{*T}
\Omega_2 (z_2-V^*_1 \vec{\epsilon}_{21} - V^*_2 \vec{\epsilon}_{22})\right),
\end{eqnarray*}
where $z_2$ is the latent variable with $z_{2j} =\frac{Y_j -r}{2\omega_{2j}}$, $ \Omega_2 = \text{diag}(\omega_{2j})$, and $V^*=[X^*, V^*_1, V^*_2]$ is the augmented design matrix for observations at risk ($W_j=1$).\\

Step 10: Update $\vec{\epsilon}_{21}$\\
Conditional on $\beta, \vec{c}, \vec{d},\vec{\epsilon}_{22}$, assuming a $\mathcal{N}(0,\sigma^2_{\epsilon_{21}})$ prior, update $\vec{\epsilon}_{21}$ from a $\mathcal{N}(\mu,\Sigma)$
distribution where
\begin{eqnarray}
  \Sigma &=&
\left(\frac{1}{\sigma^2_{\epsilon_{2S}}}I +(V^*_1)^{T}
\Omega_2 V^*_1\right)^{-1},\\
\mu &=& \Sigma
\left((V^*_1)^{T}
\Omega_2(z_2 -X \beta - V^*_1 \vec{c} -V^*_2 \vec{d} - V^*_2 \vec{\epsilon}_{22})\right).  
\label{pos_beta}
\end{eqnarray}\\

Step 11: Update $\vec{\epsilon}_{22}$\\
Conditional on $\beta, \vec{c}, \vec{d},\vec{\epsilon}_{21}$, assuming a $\mathcal{N}(0,\sigma^2_{\epsilon_{22}})$ prior, update $\vec{\epsilon}_{22}$ from a $\mathcal{N}(\mu,\Sigma)$
distribution, where
\begin{eqnarray}
  \Sigma &=&
\left(\frac{1}{\sigma^2_{\epsilon_{22}}}I +(V^*_2)^{T}
\Omega_2 V^*_2\right)^{-1},\\
\mu &=& \Sigma
\left((V^*_2)^{T}
\Omega_2(z_2 -X \beta - V^*_1 \vec{c} -V^*_2 \vec{d} - V^*_1 \vec{\epsilon}_{21})\right).  
\label{pos_beta}
\end{eqnarray}\\

Step 12: Update $l_{21}, \sigma_{21}$\\
The update of $l_{21}$ is similar to that of $l_{11}$. The Metropolis-Hasting algorithm is used to update $l_{21}$ with acceptance probability
\begin{align*}
    A & = \min \left( 1, \frac{\pi(l^*_{21})}{\pi(l_{21})} \right)\\
      & = \min \left( 1, \frac{p(\vec{c}\mid K_{\sigma_{21},l^*_{21}} (h_1, \ldots, h_S))p(l^*_{21}\mid a_{l_1} , b_{l_1})}{p(\vec{c}\mid K_{\sigma_{21},l_{21}} (h_1, \ldots, h_S))p(l_{21}\mid a_{l_1} , b_{l_1})} \right).
\end{align*}
Hyperparameter $\sigma_{21}$ is drawn from the full conditional distribution
$$
\sigma^2_{21} \sim \mathcal{IG}\left(a_{\sigma_1} + \frac{S}{2}, b_{\sigma_1} + \frac{\vec{c}^T (K_{\sigma_{21},l_{21}} (h_1, \ldots, h_S))^{-1} \vec{c}}{2}\right),
$$
where $K^{-1}_{\sigma_{21},l_{21}}$ is approximated using latent NNGP.\\

Step 13: Update $l_{22}, \sigma^2_{22}$\\
First, $l_{22}$ is updated using the Metropolis-Hasting algorithm with acceptance probability
\begin{align*}
    A & = \min \left( 1, \frac{\pi(l^*_{22})}{\pi(l_{22})} \right)\\
      & = \min \left( 1, \frac{p(\vec{d}\mid K_{\sigma_{22},l^*_{22}} (w_1, \ldots, w_T))p(l^*_{22}\mid a_{l_2} , b_{l_2})}{p(\vec{d}\mid K_{\sigma_{22},l_{22}} (w_1, \ldots, w_T))p(l_{22}\mid b_{l_2} , b_{l_2})} \right).
\end{align*}
Then, $\sigma_{22}$ can be sampled from the full conditional as
$$
\sigma^2_{22} \sim \mathcal{IG}\left(a_{\sigma_2} + \frac{T}{2}, b_{\sigma_2} + \frac{\vec{d}^T (K_{\sigma_{22},l_{22}} (w_1, \ldots, w_T))^{-1} \vec{d}}{2}\right).
$$\\

Step 14: Update $\sigma_{\epsilon_{21}}$\\
Assuming a $\mathcal{IG}(a_{\epsilon},b_{\epsilon})$ prior, draw $\sigma_{\epsilon_{21}}$ from the posterior distribution:
\begin{align*}
    \sigma^2_{\epsilon_{21}} \sim \mathcal{IG}\left(a_{\epsilon} + \frac{S}{2}, b_{ \epsilon} + \frac{\sum_{s=1}^{S}\epsilon^2_{21s}}{2}\right).
\end{align*}\\

Step 15: Update $\sigma_{\epsilon_{22}}$\\
Assuming a $\mathcal{IG}(a_{\epsilon},b_{\epsilon})$ prior, draw $\sigma_{\epsilon_{22}}$ from the posterior distribution:
\begin{align*}
    \sigma^2_{\epsilon_{22}} \sim \mathcal{IG}\left(a_{\epsilon} + \frac{T}{2}, b_{ \epsilon} + \frac{\sum_{t=1}^{T}\epsilon^2_{22t}}{2}\right).
\end{align*}

\bibliography{main_JSPI}

\end{document}